\documentclass[10pt,twocolumn,letterpaper]{article}

\usepackage[
  letterpaper,
  top=0.72in,
  bottom=0.82in,
  left=0.72in,
  right=0.72in,
  columnsep=0.24in
]{geometry}
\usepackage[numbers,sort&compress]{natbib}
\usepackage[T1]{fontenc}
\usepackage[utf8]{inputenc}
\usepackage{microtype}
\usepackage{mathptmx}
\usepackage{amsmath}
\usepackage{amssymb}

\usepackage{graphicx}
\usepackage[table]{xcolor}
\usepackage{booktabs}
\usepackage{multirow}
\usepackage{caption}
\usepackage{newfloat}

\usepackage{algorithm}
\usepackage{algorithmic}
\usepackage{listings}
\DeclareCaptionStyle{ruled}{labelfont=normalfont,labelsep=colon,strut=off}
\floatstyle{ruled}
\newfloat{listing}{tb}{lst}{}
\floatname{listing}{Listing}

\usepackage[many]{tcolorbox}
\newtcolorbox{promptbox}[1][]{
  colback=gray!5!white,
  colframe=gray!75!black,
  fonttitle=\bfseries,
  title=Prompt,
  breakable,
  boxsep=2pt,
  left=4pt,
  right=4pt,
  top=4pt,
  bottom=4pt,
  #1
}

\usepackage[hyphens]{url}
\usepackage[
    colorlinks=true,
    linkcolor=black,
    citecolor=black,
    urlcolor=blue
]{hyperref}
\usepackage{footmisc}
\title{SonicWeave: Chunk-Routed Mixture-of-Experts for Unified \\Audio Scene Generation}
\author{
  Yunrui Cai\textsuperscript{1,2,*}, \
  Xu Li\textsuperscript{1,$\dagger$}, \
  Yucheng Zhou\textsuperscript{1}, \
  Jinchao Li\textsuperscript{1}, \
  Dingdong Wang\textsuperscript{2}, \
  Dongchao Yang\textsuperscript{2}, \\
  Xixin Wu\textsuperscript{2}, \
  Chen Zhang\textsuperscript{1}, \
  Zhiyong Wu\textsuperscript{3}, \
  Pengfei Wan\textsuperscript{1}, \
  Helen Meng\textsuperscript{2,$\dagger$}\\[0.45em]
  \small \textsuperscript{1}Kling Team, Kuaishou Technology \\
  \small \textsuperscript{2}The Chinese University of Hong Kong\\
  \small \textsuperscript{3}Shenzhen International Graduate School, Tsinghua University\\
  \small \texttt{\{yrcai, hmmeng\}@se.cuhk.edu.hk, lixu15@kuaishou.com}
}
\date{}

\begin{document}

% Full-width title and abstract above the two-column body.
\makeatletter
\twocolumn[
\begin{@twocolumnfalse}
\maketitle
\begin{abstract}
Text-conditioned general audio generation is moving beyond isolated speech, music, and sound-effect synthesis toward a single model that can compose them into controllable, coherent audio scenes. This unified setting is particularly challenging: heterogeneous components impose conflicting structural requirements on a shared backbone, while a complex mixed scene may contain locally distinct or overlapping content that demands fine-grained adaptation within the same clip. Existing audio mixture-of-experts (MoEs) mainly route at the domain level, while token-wise routing overlooks the local continuity inherent to acoustic signals. We propose SonicWeave, a flow-matching model for unified audio scene generation. At its core is a chunk-routed MoE with a conflict-gated prior–evidence routing mechanism (CPE-MoE). CPE-MoE routes contiguous acoustic chunks by combining a global prior that encodes the structured text condition and diffusion phase with local evidence from the evolving acoustic state. A learned conflict gate favors the prior when local states are unreliable, while allowing local evidence to influence routing when a region departs from the global scene context. SonicWeave supports speech, music, sound effects, singing, and their fine-grained mixtures with a single set of weights. 
% Across TTS, TTA, and TTM benchmarks, SonicWeave achieves strong generation quality and consistently improves over controlled dense and base-MoE baselines. Evaluation on complex mixed scenes further confirms improved compositional quality, while routing analyses reveal content-dependent expert specialization and systematic adaptation across diffusion phases.
Across TTS, TTA, and TTM benchmarks, SonicWeave consistently improves over controlled Dense and Base-MoE baselines. Complex-scene evaluation further demonstrates improved compositional quality, while routing analyses reveal content-dependent expert specialization across diffusion phases. These results suggest that temporally coherent, prior–evidence routing is an effective conditional-computation strategy for unified audio generation. Project page: \url{https://caiyunrui.github.io/SonicWeave}.
\end{abstract}
\vspace{0.8em}
\end{@twocolumnfalse}
]
\makeatother
\begingroup
\renewcommand{\thefootnote}{\fnsymbol{footnote}}
\footnotetext[1]{Work done during an internship at Kling Team, Kuaishou Technology.}
\footnotetext[2]{Corresponding authors.}
\endgroup

\section{Introduction}

\begin{figure*}[t]
\centering
% Full-width teaser for the arXiv layout.
\includegraphics[width=0.62\textwidth]{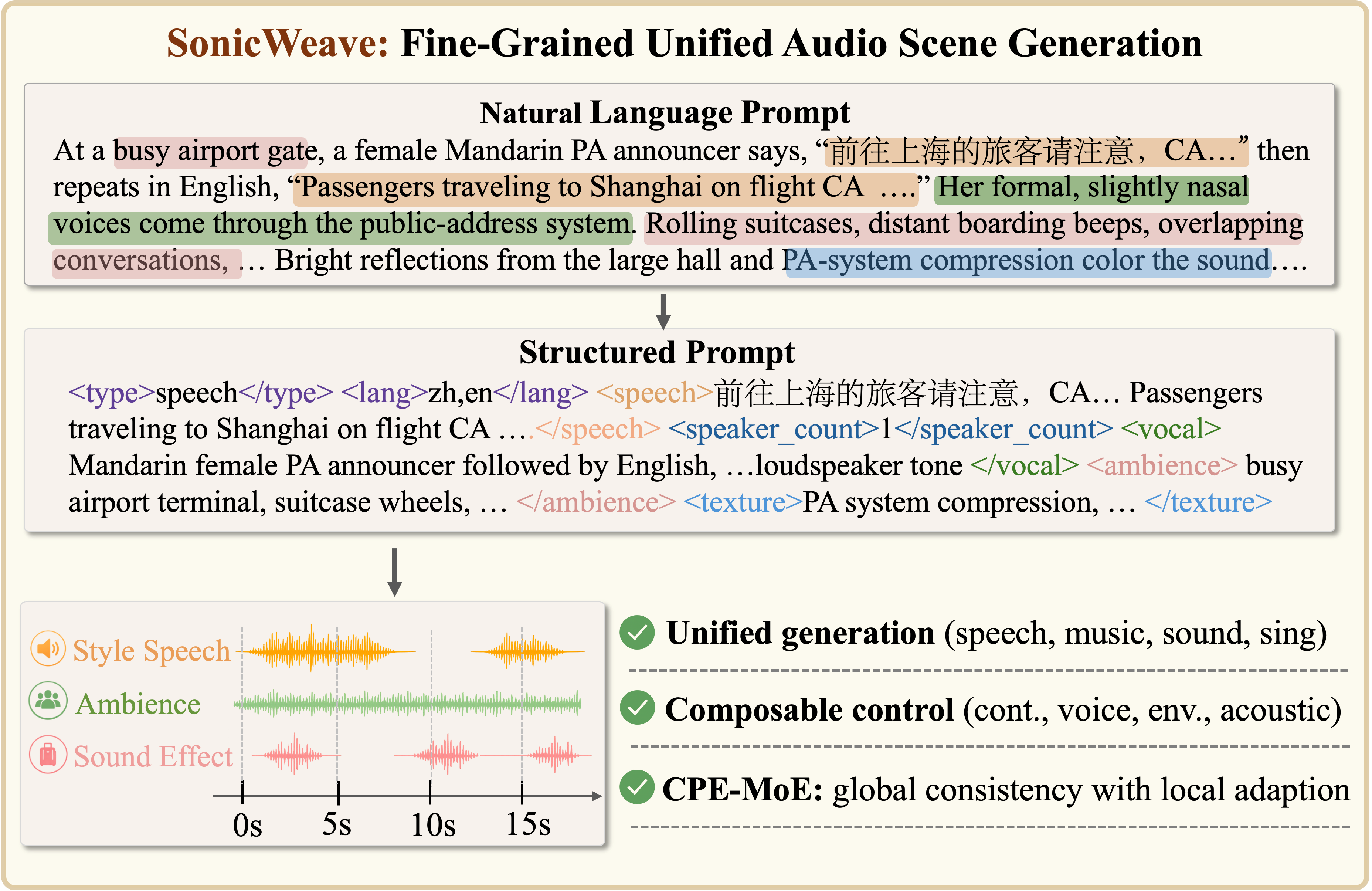}
\caption{Overview of SonicWeave for unified audio scene generation. Natural-language descriptions are converted into structured captions that separate speech content, vocal attributes, music, sound effects, and ambience. A single model then synthesizes these heterogeneous components jointly, while its CPE-MoE module provides conflict-gated, content-aware routing across acoustic chunks.}
\label{fig:teaser}
\end{figure*}

Recent diffusion and flow-matching models~\cite{flow} have broadened text-conditioned synthesis across speech, music, and environmental sounds \cite{voicebox, f5tts, tangoflux}. The next step is \emph{general audio generation}: one model should not only cover these domains, but also compose them into coherent scenes and respond to structured specifications of heterogeneous acoustic content---for example, narration combined with music, ambience, and transient sound events \cite{audioldm2, uniaudio, audiobox, dashengaudiogen}. This capability is important for moving from isolated audio generation to controllable modeling of real acoustic environments.

This goal exposes a fundamental conditional-computation problem at two granularities: across domains (speech vs. music) and within a scene (locally overlapping components). Across domains, speech requires linguistic precision, music requires harmonic and rhythmic structure, and environmental sound includes both stationary textures and transient events. Within a single complex scene, these components may overlap or alternate, so different local regions require different computation and fine-grained control of heterogeneous content. A dense diffusion transformer (DiT) applies the same feed-forward transformation to every acoustic token, while existing audio MoEs primarily target domain-level specialization \cite{unimoeaudio}. Token-wise routing may also vary unnecessarily across correlated neighboring frames, and local acoustic states are unreliable at early noisy diffusion phases. A useful unified generator therefore needs to coordinate global scene intent, local acoustic evidence, and temporal continuity rather than treating routing as an independent decision for every token.

Acoustic latents exhibit strong short-range dependence: neighboring frames typically belong to the same phonetic unit, musical texture, or environmental event, even when the broader scene contains heterogeneous components. Chunk-level routing therefore occupies a useful middle ground between clip-level routing, which cannot adapt within a scene, and frame-level routing, which may fragment locally coherent structure.

We propose SonicWeave, a unified flow-matching model designed for complex-scene generation with fine-grained local adaptation. Its core CPE-MoE module---Conflict-gated Prior--Evidence Mixture-of-Experts---routes contiguous acoustic chunks rather than independent frames. For each chunk, CPE-MoE combines a global prior from structured scene text and diffusion phase with local evidence from the evolving acoustic representation. 
The learned conflict gate favors the prior when local states are unreliable and shifts toward local evidence when a region contains content not explained by the global scene context. 
% The learned conflict gate favors the prior when local states are unreliable and permits greater influence from local evidence when the evolving chunk representation conflicts with, or departs from, the global prior.
Text and time tokens remain on a shared expert, preserving a stable conditioning pathway. SonicWeave is trained over speech, music, sound effects, singing, and mixed scenes under one structured textual interface.

Experiments show that SonicWeave performs strongly across TTS, TTA, and TTM while consistently improving over controlled Dense and Base-MoE baselines. On controlled complex scenes, it produces more coherent mixtures and better follows structured specifications of foreground, background, and speaker or vocal attributes. Routing analyses further show content-dependent expert specialization and diffusion-phase-adaptive use of the global prior and local evidence.

Our contributions are threefold:

\begin{itemize}
% \item We formulate unified audio scene generation as a joint challenge of broad domain coverage, complex-scene composition, and structured control of heterogeneous acoustic content, using a textual interface for speech, music, sound effects, and ambience.
\item We identify conditional computation at two coupled granularities—across audio domains and within heterogeneous scenes—and formulate routing requirements that jointly account for global scene intent, local acoustic state, and short-range temporal continuity.
% \item We introduce CPE-MoE, an audio-aware chunk routing module that provides stable local specialization by reconciling a text-and-phase prior with evolving acoustic evidence through a learned conflict gate.
\item We introduce CPE-MoE, which routes contiguous acoustic chunks using a diffusion-phase-aware text prior, local acoustic evidence, and a learned conflict gate, while keeping text and time tokens on a stable shared path.
% \item We demonstrate strong performance across TTS, TTA, and TTM, improved compositional quality on complex mixed scenes, and interpretable routing behavior that reveals content- and phase-dependent expert specialization.
\item Through matched Dense and token-routed MoE controls, public-task benchmarks, a 100-prompt complex-scene suite, human listening tests, and routing analyses, we show that the proposed routing improves compositional fidelity without sacrificing perceptual quality.
\end{itemize}

\section{Related Work}

\begin{figure*}[t]
\centering
% TODO: Figure 2: left, dual-stream audio--text flow-matching DiT; right, CPE-MoE routing block.
\includegraphics[width=\linewidth]{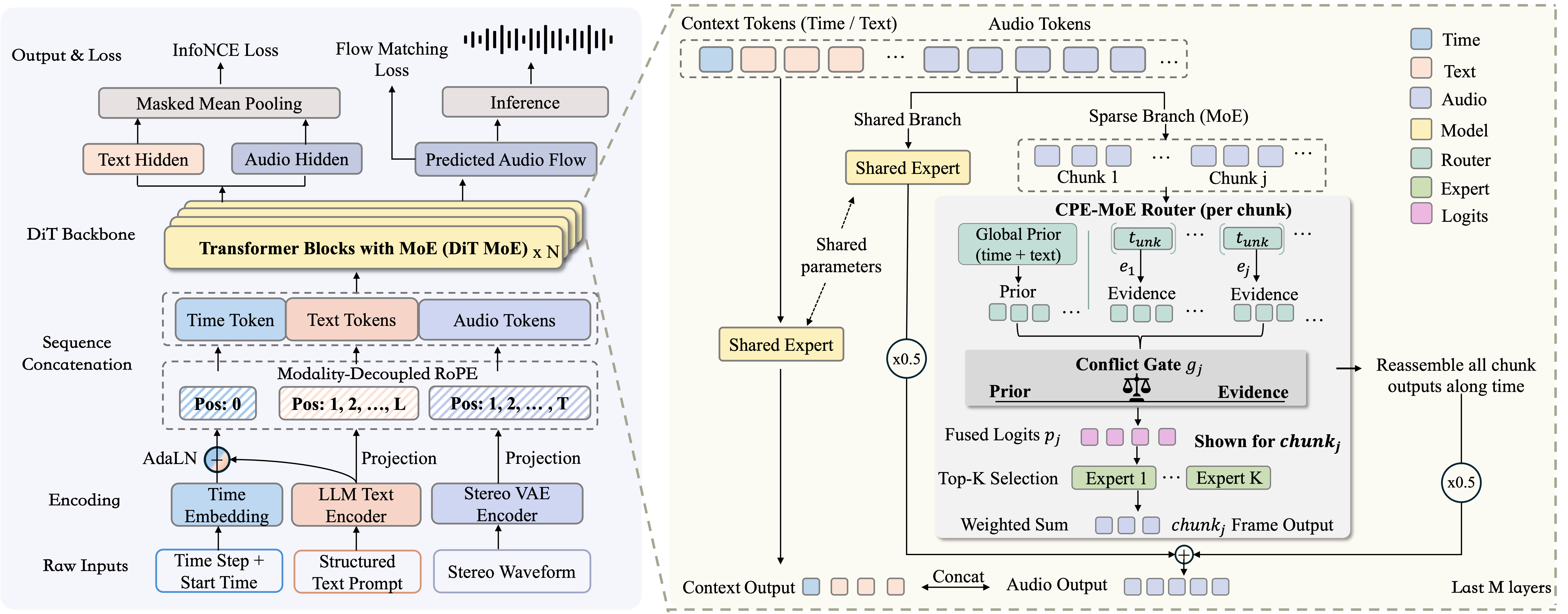}
\caption{SonicWeave framework. \textbf{Left}: a stereo VAE encodes audio into continuous latents, which are jointly modeled with a structured-caption stream in a flow-matching DiT. \textbf{Right}: the CPE-MoE module groups audio tokens into chunks and fuses a global text-and-phase prior with local acoustic evidence for sparse expert routing.}
\label{fig:framework}
\end{figure*}

\subsection{General and Unified Audio Generation}

Text-conditioned audio generation has progressed from autoregressive audio language models to diffusion and flow-matching systems for speech, music, and environmental sounds \cite{audioldm, voicebox, f5tts, tangoflux}. Unified models extend this scope through shared representations or architectures \cite{audioldm2, uniaudio, audiobox, unisonate}. Dasheng AudioGen further targets coherent mixed scenes with structured multi-view captions \cite{dashengaudiogen}. 
% In contrast, we study conditional computation inside a shared continuous generator: routing must address not only broad domain differences but also local conflicts among coexisting components within a scene.
In contrast, we focus on conditional computation within a unified continuous generator, where routing must resolve both heterogeneous audio domains and local compositional conflicts among coexisting scene components.

\subsection{Structured Text Conditioning}

Structured conditions improve controllability by exposing event, temporal, or semantic factors beyond a single free-form caption \cite{makeanaudio2, tango, freeaudio}. Our structured caption normalizes heterogeneous annotations into one interface for speech content and attributes, music, sound events, and ambience. Unlike explicit timeline planning, it supplies a common scene description; CPE-MoE, the core module of SonicWeave, uses this global information as a routing prior and combines it with evolving local acoustic evidence.

\subsection{Mixture-of-Experts for Diffusion and Audio Generation}

MoE enables conditional capacity through sparse expert activation and has been adapted to diffusion transformers \cite{switchtransformer, ditmoe, switchdit, ecdit}. In unified audio generation, UniMoE-Audio uses specialist-initialized experts and dynamic allocation to mitigate speech--music task conflict \cite{unimoeaudio}. These approaches establish the value of specialization, but operate at task/modality granularity or route individual tokens. SonicWeave instead uses CPE-MoE to route contiguous acoustic chunks with a conflict gate between a global text-and-phase prior and local evidence, targeting stable, fine-grained specialization in heterogeneous mixed scenes.

\section{Method}

\subsection{Overview}
Our dual-stream conditional flow-matching generator, SonicWeave, jointly models audio and text. In the audio stream, a stereo waveform is encoded by a pretrained stereo VAE into continuous audio latents; the model denoises a noised latent together with an optional masked reference for continuation and inpainting. In the text stream, a frozen language encoder embeds a structured audio caption. A joint audio--text diffusion transformer (DiT) then predicts the flow velocity, with the CPE-MoE module replacing the FFN in its final layers. Figure~\ref{fig:framework} summarizes the SonicWeave architecture and its CPE-MoE block.

\subsection{Structured Text Conditioning}
We convert each natural-language annotation into a compact, audio-only structured caption. An automatic annotation pipeline first removes non-acoustic details and extracts a fixed set of fields: type, language, speech text or lyrics, speaker counts, speaker and vocal attributes, music, sound effect, ambience, and recording texture. The fields are serialized with explicit tags, e.g., \texttt{Summary sentence. <type>speech</type> <lang>en</lang> <speech>...</speech> <music>...</music>}. 

This factorized format separates otherwise entangled scene attributes, preserves linguistic content verbatim, and supplies a consistent interface across speech, music, sound effects, and mixed audio. The current schema explicitly supports up to two speakers for dialogue-oriented scenes. A frozen language encoder maps the resulting caption into text tokens for the DiT; we use one shared encoder rather than domain-specific conditioning modules.

\subsection{Joint Audio--Text DiT Backbone}

\subsubsection{Sequence Construction.}
Each transformer layer operates on the concatenation of three modality groups:
\begin{equation}
\mathbf{h}=\big[\underbrace{\mathbf{t}_{\text{time}}}_{1}\;\|\;\underbrace{\mathbf{T}_{1:L}}_{\text{text}}\;\|\;\underbrace{\mathbf{A}_{1:T}}_{\text{audio}}\big]\in\mathbb{R}^{(1+L+T)\times d}.
\end{equation}
The time token $\mathbf{t}_{\text{time}}$ is obtained by summing sinusoidal embeddings for diffusion time $t$ and segment start $s$, then projecting the result to $d$. Text tokens are $\mathbf{T}=W_{\text{cap}}\mathbf{E}$, where $\mathbf{E}$ denotes the frozen language-encoder outputs. Audio tokens concatenate the noised stereo-VAE latent $\mathbf{x}_t$ with the masked reference $\mathbf{c}$, followed by linear projection and convolutional positional embedding.

\subsubsection{Modality-Decoupled RoPE.}
Inspired by the modality-specific coordinate design of M-RoPE \cite{qwen2vl}, we assign text and audio independent one-dimensional position axes. The time token uses position $0$, while text and audio positions are reset separately:
\begin{equation}
\operatorname{pos}(\mathbf{t}_{\mathrm{time}})=0, \ 
\operatorname{pos}(\mathbf{T}_i)=i,\ 
\operatorname{pos}(\mathbf{A}_j)=j.
\end{equation}
With this modality-specific coordinate system, positional encoding preserves the intrinsic order of both text and audio while keeping their coordinates independent of sequence length in the other modality. It also provides a consistent relative-position structure for joint attention to model cross-modal correspondences.

\subsubsection{Text-Conditioned AdaLN.}
Joint self-attention provides token-level text--audio interaction, while adaptive normalization offers a complementary path for injecting global conditions into diffusion transformers \cite{dit}. 
% We derive this global condition through generation-phase-aware attention pooling. 
We additionally use phase-aware attention pooling over the structured
caption to provide global AdaLN conditioning; details are given in the
appendix.
% Let $\mathbf{u}=\mathbf{t}+\mathbf{s}$ combine the diffusion-time and segment-start embeddings, and let $\mathbf{E}\in\mathbb{R}^{L\times d_E}$ denote the frozen text features. We compute
% \begin{equation}
% \begin{aligned}
% \mathbf{q} &= W_q\mathbf{u}, \ \mathbf{K} = W_k\mathbf{E}, \ \mathbf{v} = \boldsymbol{\alpha}\mathbf{E}, \\
% \boldsymbol{\alpha} &= \operatorname{softmax}\!\left(
% \mathbf{q}\mathbf{K}^{\top}/\sqrt{d_E}+\mathbf{M}\right),\
% \mathbf{c} = \mathbf{u}+W_{\mathrm{global}}\mathbf{v}.
% \end{aligned}
% \end{equation}
% Here $\mathbf{M}$ masks padded text positions, and $\mathbf{c}$ modulates the final AdaLN-Zero layer. Unlike static pooling, the phase-derived query $\mathbf{q}$ lets the model attend to different caption fields over the generation trajectory, while the residual $\mathbf{u}$ retains explicit diffusion-phase information.

\subsubsection{Cross-Modal Alignment Loss.}
The shared DiT processes audio and text jointly, but flow matching alone does not explicitly require their final hidden states to be aligned. We therefore add a symmetric in-batch contrastive objective, following the paired-representation alignment principle used in CLIP \cite{clip}. Let $\mathbf{a}_b$ and $\mathbf{r}_b$ be the $\ell_2$-normalized, mask-pooled audio and text states of sample $b$, respectively, and define $S_{b b'}=\exp(\lambda)\,\mathbf{a}_b^\top\mathbf{r}_{b'}$, where $\lambda$ is learned. We minimize
\begin{equation}
\mathcal{L}_{\mathrm{CL}}=\tfrac12\left[\operatorname{CE}(\mathbf{S},\mathbf{y})+\operatorname{CE}(\mathbf{S}^{\top},\mathbf{y})\right],\ y_b=b.
\end{equation}
This aligns paired audio--caption representations and supports the text-derived prior and audio-derived evidence used by CPE-MoE.
% The loss aligns each audio scene with its paired caption and separates it from other batch examples. This encourages text--audio correspondence within the shared backbone and supports semantically compatible representations for the text-derived prior and audio-derived evidence used by CPE-MoE.
% We omit this term for caption-dropped or single-sample batches.

\subsection{CPE-MoE: Conflict-Gated Prior--Evidence Chunk Routing}
The CPE-MoE module replaces the dense FFN in the final transformer layers with shared and sparsely routed experts, providing conditional capacity for heterogeneous audio. As shown in Figure~\ref{fig:framework}, it learns routing from two complementary sources: structured text and diffusion phase specify global generation intent, while evolving audio states reveal local acoustic content. Routing contiguous chunks preserves short-range audio continuity, allowing expert specialization across broad tasks and locally mixed content under one objective.

\subsubsection{Modality-Decoupled Dispatch.}
Given the layer input $\mathbf{h}\!\in\!\mathbb{R}^{B\times(1+L+T)\times d}$, we split it into a context group (time and text tokens) and an audio group:
\begin{equation}
\mathbf{h}_{\text{ctx}}=\mathbf{h}_{[:,\,:1{+}L]},\
\mathbf{h}_{\text{aud}}=\mathbf{h}_{[:,\,-T:]}.
\end{equation}
The context group is processed only by the shared expert $\mathbf{o}_{\text{ctx}}=\mathrm{MLP}_{\text{shared}}(\mathbf{h}_{\text{ctx}})$. This isolates the text and time representations from the routing gate.

\subsubsection{Chunk-Level Routing.}
We partition the $T$ audio states into $N_c=\lceil T/C\rceil$ non-overlapping chunks $\{\mathcal{C}_j\}_{j=1}^{N_c}$ of at most $C$ consecutive frames. Let $m_i\in\{0,1\}$ indicate whether frame $i$ is valid. The pooled state of chunk $j$ is
\begin{equation}
\mathbf{e}_j=
\frac{\sum_{i\in\mathcal{C}_j}m_i\mathbf{h}_{\mathrm{aud},i}}
     {\max\!\left(1,\sum_{i\in\mathcal{C}_j}m_i\right)} .
\end{equation}
Padding-only chunks are excluded from routing and auxiliary statistics. The prior--evidence router makes one top-$K$ assignment per valid chunk and broadcasts it to all frames in that chunk. The selected experts nevertheless transform the original frame states, not the pooled state $\mathbf{e}_j$. Thus, pooling controls only expert selection: frame-level modeling is retained while neighboring frames share a locally consistent computation path.
Importantly, the chunk is the routing unit rather than the representation unit: expert selection is shared within a chunk, but each expert still transforms the original frame-level states.

\subsubsection{Prior and Evidence Signals.}
The router combines two complementary signals with different information scopes for each chunk.

\noindent\textbf{Prior.} A global, chunk-independent signal that summarizes \emph{what to generate} and \emph{at which diffusion phase}:
\begin{equation}
\mathbf{p}=[\mathbf{t}_{\text{time}}\;\|\;\bar{\mathbf{T}}]\in\mathbb{R}^{2d},\
\boldsymbol{\ell}^{\mathrm{prior}}=W_{\mathrm{prior}}(\mathbf{p})\in\mathbb{R}^{N_e},
\end{equation}
where $\bar{\mathbf{T}}$ is the mask-pooled text hidden state at the current layer. Because $\mathbf{p}$ carries both the text summary and the diffusion-step embedding, $\boldsymbol{\ell}^{\mathrm{prior}}$ is a task- and step-aware expert preference that summarizes the global generation intent.

\noindent\textbf{Evidence.} A local signal is computed from the current context-conditioned audio representation $\mathbf{e}_j$:
\begin{equation}
\boldsymbol{\ell}^{\mathrm{evid}}_j=W_{\mathrm{evid}}(\mathbf{e}_j)\in\mathbb{R}^{N_e}.
\end{equation}
Because $\mathbf{e}_j$ is pooled after joint self-attention, it reflects the acoustic state currently emerging in chunk $j$. The evidence logits can thus capture local departures from the scene-level description, such as a transient event within speech or music.

Both projections are 2-layer MLPs with SiLU activation and zero-initialized output biases, avoiding an explicit expert preference at initialization.

\subsubsection{Conflict Gate and Fusion.}

\begin{table*}[t]
\centering
\caption{Comparison with representative specialized and general-purpose audio generators. $\checkmark$ indicates explicitly reported support; $\triangle$  denotes partial support, limited public demonstrations or support under restricted settings; and -- denotes no explicit evidence of support found in the public documentation we reviewed. "Background" includes music, sound effects, environmental ambience, or their combinations.}
\label{tab:model_scope}
\resizebox{\textwidth}{!}{%
\begin{tabular}{l|l|ccc|cccc}
\toprule
\multirow{2}{*}{Category} & \multirow{2}{*}{Model}  & \multicolumn{3}{c|}{General tasks} & \multicolumn{4}{c}{Complex-scene generation} \\
\cmidrule(lr){3-5}\cmidrule(lr){6-9}
&  & TTS & TTA & TTM & Speech+Bg. & Sing+Bg. & Dialog & Dialog+Bg. \\
\midrule
\multirow{6}{*}{Specialized}
& CosyVoice 2~\cite{cosyvoice2}           & $\checkmark$ & -- & -- & -- & -- & -- & -- \\
& F5-TTS~\cite{f5tts}                  & $\checkmark$ & -- & -- & -- & -- & -- & -- \\
& Step-Audio~\cite{stepaudio}                       & $\checkmark$ & -- & -- & -- & -- & $\triangle$ & -- \\
& TangoFlux~\cite{tangoflux}                    & -- & $\checkmark$ & -- & -- & -- & -- & -- \\
& MusicGen-Large~\cite{musicgen}        & -- & -- & $\checkmark$ & -- & -- & -- & --  \\
& SegTune~\cite{segtune}        & -- & -- & $\checkmark$ & -- & $\triangle$ & -- & --  \\
\midrule
\multirow{9}{*}{General / multi-task}
& Qwen2.5-Omni~\cite{qwen25omni}                & $\checkmark$ & $\triangle$ & $\triangle$ & -- & -- & $\triangle$ & -- \\
& AudioLDM~2~\cite{audioldm2}            & $\triangle$ & $\checkmark$ & $\triangle$ & $\triangle$ & $\triangle$ & -- & -- \\
& Stable Audio Open~\cite{stableaudio}  & -- & $\checkmark$ & $\checkmark$ & -- & $\triangle$ & -- & -- \\
& UniAudio~\cite{uniaudio}               & $\checkmark$ & $\checkmark$ & $\checkmark$ & $\triangle$ & -- & -- & -- \\
& AudioX~\cite{audiox}                        & -- & $\checkmark$ & $\checkmark$ & -- & -- & -- & -- \\
& Higgs Audio V2~\cite{higgsaudio}                 & $\checkmark$ & $\checkmark$ & $\triangle$ & $\triangle$ & -- & $\checkmark$ & $\triangle$ \\
& UniMoE-Audio~\cite{unimoeaudio}        & $\checkmark$ & -- & $\checkmark$ & -- & -- & -- & -- \\
& UniFlow-Audio~\cite{uniflowaudio}    & $\checkmark$ & $\checkmark$ & $\checkmark$ & $\triangle$ & -- & $\triangle$ & $\triangle$ \\
& Dasheng AudioGen~\cite{dashengaudiogen}  & $\checkmark$ & $\checkmark$ & $\checkmark$ & $\checkmark$ & -- & $\triangle$ & $\triangle$ \\
% & Dense (ours)                           & 2.0B  & $\checkmark$ & $\checkmark$ & $\checkmark$ & $\checkmark$ & $\checkmark$ & $\checkmark$ & $\checkmark$ \\
% & BaseMoE (ours)                         & 2.0B  & $\checkmark$ & $\checkmark$ & $\checkmark$ & $\checkmark$ & $\checkmark$ & $\checkmark$ & $\checkmark$ \\
\midrule
% \rowcolor{cyan!9}
\textbf{General / compositional} & \textbf{SonicWeave (ours)}  & $\checkmark$ & $\checkmark$ & $\checkmark$ & $\checkmark$ & $\checkmark$ & $\checkmark$ & $\checkmark$ \\
\bottomrule
\end{tabular}%
}
\end{table*}

The prior gives a global expert preference, but local audio becomes informative at different rates across diffusion phases and acoustic regions. We therefore let each chunk decide how far its routing should move from the prior toward the local evidence. We first project the semantic--phase summary into the $d$-dimensional hidden space of the joint DiT: 
\begin{equation}
\tilde{\mathbf{p}}=W_{\mathrm{align}}\mathbf{p}\in\mathbb{R}^{d}.
\end{equation}
% This does not compare raw features from separate encoders: $\mathbf{p}$ and $\mathbf{e}_j$ are formed from the same post-attention DiT layer, and $W_{\mathrm{align}}$ is learned jointly to make their coordinates comparable.

The gate observes both signals and their explicit pairwise interactions:
\begin{equation}
\boldsymbol{\phi}_j=\mathrm{LN}\!\left(
[\,\tilde{\mathbf{p}}\;\|\;\mathbf{e}_j\;\|\;
\tilde{\mathbf{p}}\!\odot\!\mathbf{e}_j\;\|\;
|\tilde{\mathbf{p}}-\mathbf{e}_j|\,]\right),
\end{equation}  
\begin{equation}
g_j=\sigma\!\left(\operatorname{MLP}_{g}(\boldsymbol{\phi}_j)\right).
\end{equation}
Following pairwise matching features \cite{infersent}, the product exposes coordinate-wise agreement and the absolute difference exposes disagreement, while the original vectors retain source-specific information. The MLP therefore learns how agreement or conflict should affect routing rather than imposing a fixed similarity rule; this is useful because discrepancy may reflect either a meaningful local event or noisy evidence depending on diffusion phase. The gate is trained by the generation objective as a conflict-conditioned evidence-reliance coefficient, not as a calibrated probability of evidence correctness.

The meaning of $g_j$ acts directly in the expert-logit space: the prior and evidence routers first produce logits of the same shape, and $g_j$ linearly interpolates between them:
\begin{equation}
\boldsymbol{\ell}_j=(1-g_j)\boldsymbol{\ell}^{\mathrm{prior}}
+g_j\boldsymbol{\ell}^{\mathrm{evid}}_j,
\ 
\mathbf{P}_j=\operatorname{softmax}(\boldsymbol{\ell}_j).
\end{equation}
Thus, $g_j=0$ exactly recovers the prior logits, whereas $g_j=1$ exactly recovers the evidence logits. 
% More generally,
% $\partial\boldsymbol{\ell}_j/\partial g_j=
% \boldsymbol{\ell}^{\mathrm{evid}}_j-\boldsymbol{\ell}^{\mathrm{prior}}$;
% training therefore increases $g_j$ when shifting the routing decision toward local evidence helps reduce the generation objective, and decreases it when the prior is more useful. We consequently interpret $g_j$ as a \emph{conflict-conditioned evidence-reliance coefficient}, rather than a calibrated probability of evidence correctness. 
The top-$K$ entries of $\mathbf{P}_j$ select the experts, and their renormalized probabilities directly weight the expert outputs.

\subsubsection{Sparse Dispatch and Composition.}
% Top-$K$ selection on $\mathbf{P}_j$ yields per-chunk expert indices $\{e_j^k\}_{k=1}^K$ and re-normalised weights $\{w_j^k\}$. These are broadcast to the $C$ frames of chunk $j$. For each expert we gather the active chunks, apply the expert MLP, scale by $w_j^k$, and accumulate into the sparse output; padding frames inside a chunk are masked out.

Top-$K$ indices and weights are broadcast to frames within each chunk; active expert outputs are weighted and accumulated, with padding masked out.

The audio output of the layer is an \emph{equal-weight} combination of the shared expert and the sparse mixture,
\begin{equation}
\mathbf{o}_{\text{aud}}=\tfrac12\,\mathrm{MLP}_{\text{shared}}(\mathbf{h}_{\text{aud}})+\tfrac12\,\mathbf{o}_{\text{sparse}}.
\end{equation}
The down-projection of each routed expert is zero-initialized, so the sparse contribution starts near zero and is introduced smoothly during optimization. The fixed $\tfrac12{:}\tfrac12$ composition prevents an additional learnable mixture gate from suppressing the routed path; gradients can update the expert down-projections from the first step, after which the experts and router co-adapt. The final layer output is $[\mathbf{o}_{\text{ctx}}\;\|\;\mathbf{o}_{\text{aud}}]$.

\subsubsection{Load Balancing.}
To prevent expert collapse we adopt a Switch-Transformer-style auxiliary loss~\cite{switchtransformer}, computed at chunk granularity and only over valid chunks:
\begin{equation}
\mathcal{L}_{\mathrm{MoE}}=N_e\cdot\sum_{i=1}^{N_e}\bar{\mathbf{p}}_i^{\text{valid}}\cdot\bar{\mathbf{d}}_i^{\text{valid}},
\end{equation}
where $\bar{\mathbf{p}}^{\text{valid}}$ is the mean router probability over valid chunks and $\bar{\mathbf{d}}^{\text{valid}}$ is the mean top-$K$-normalised dispatch fraction. Padded chunks are excluded from both statistics. The layer-wise auxiliary losses are summed across the $M$ CPE-MoE layers and weighted in the composite objective below.

\subsection{Conditional Flow-Matching Training}

Given the ground-truth stereo-VAE latent $\mathbf{x}_1$, we sample $\mathbf{x}_0\!\sim\!\mathcal{N}(0,I)$ and $t\!\sim\!\sigma(\mathcal{N}(0,1))$, form $\mathbf{x}_t{=}(1{-}t)\mathbf{x}_0{+}t\mathbf{x}_1$, and regress the ground-truth velocity:
\begin{equation}
% \mathcal{L}_{\mathrm{FM}}=\mathbb{E}\big\|\mathbf{v}_\theta(\mathbf{x}_t,\mathbf{c}_{\text{text}},t)-(\mathbf{x}_1-\mathbf{x}_0)\big\|^2. \\
\mathcal{L}_{\mathrm{FM}}=\mathbb{E}\big\|\mathbf{v}_\theta(\mathbf{x}_t,\mathbf{c},\mathbf{c}_{\text{text}},t)-(\mathbf{x}_1-\mathbf{x}_0)\big\|^2.
\end{equation}

% \noindent\textbf{CFG dropout.} During training, the audio reference is independently dropped with probability $p_{\mathrm{ref}}$ and the caption with probability $p_{\mathrm{cap}}$. Caption drop zeros the text-token sequence and removes text from the AdaLN conditioning vector.
% \noindent\textbf{Inpainting mask.} A fractional span with ratio uniform in $[0.7,1.0]$ is chosen per sample to mark the ``to-generate'' region; those frames are zeroed in the reference $\mathbf{c}$ but preserved in the target $\mathbf{x}_1$. 
During training, we randomly mask a contiguous span of the audio reference and compute the reconstruction loss on the masked region; the audio reference and caption are independently dropped for classifier-free guidance~\cite{cfg}. 
% This strategy exposes the model to varying conditioning patterns and helps the flow-matching objective converge more quickly and stably.

The composite objective is
\begin{equation}
\mathcal{L}=\mathcal{L}_{\mathrm{FM}}+\alpha_{\mathrm{CL}}\mathcal{L}_{\mathrm{CL}}+\alpha_{\mathrm{MoE}}\mathcal{L}_{\mathrm{MoE}}.
\end{equation}

\subsection{Inference}
We integrate the learned velocity field with a fixed-step Euler ODE solver, using $100$ uniform steps in $t\!\in\![0,1]$; an optional cosine sway sampling schedule accelerates inference at low step counts.

At inference we adopt Adaptive Projected Guidance (APG)~\cite{apg} and adapt it to the clean-sample estimate induced by the CFM velocity parameterisation. With conditional and unconditional clean-sample estimates $\hat{\mu}$ and $\hat{\mu}_{\varnothing}$, let $\Delta\mu=\hat{\mu}-\hat{\mu}_{\varnothing}$ and decompose it into components parallel and orthogonal to $\hat{\mu}$, denoted by $\Delta\mu_{\parallel}$ and $\Delta\mu_{\perp}$. The guided estimate is
\begin{equation}
\hat{\mu}_{\mathrm{APG}}=\hat{\mu}_{\varnothing}+w\left(\eta\,\Delta\mu_{\parallel}+\Delta\mu_{\perp}\right),
\end{equation}
where $w$ is the guidance scale and $\eta$ damps the parallel component. It is then converted back to velocity space for ODE integration. Momentum smoothing, the clean-sample conversion, and the complete update are given in the appendix. The final clean stereo latent is decoded into waveform audio by the pretrained stereo VAE decoder.

\section{Experiments}
\subsection{Experimental Setup}

\subsubsection{Tasks and Benchmarks.}
We evaluate one unified model on three public generation tasks and a controlled complex-scene set. For text-to-speech (TTS), we use SeedTTS-eval~\cite{seedtts} and LibriSpeech-PC test-clean~\cite{librispeechpc}; for text-to-audio (TTA), we use AudioCaps~\cite{audiocaps}; for text-to-music (TTM), we use MusicCaps~\cite{musiccaps} and Song Describer~\cite{songdescriber}. The Complex-Scene set is described in the subsequent section.

For SonicWeave and controlled baselines, 
% we convert each benchmark's original transcript or natural-language caption into the structured format using rule-based templates, without introducing additional acoustic descriptions; conversion details are provided in the appendix.
we deterministically render benchmark inputs into a fixed schema using constrained field-extraction rules. Where automatic parsing is needed, a zero-temperature language model acts only as a schema-aware extractor and validator; it is prohibited from introducing information absent from the original input, apart from fixed template defaults applied uniformly across the benchmark. The converter does not observe reference audio.

\subsubsection{Baselines.}
% We compare against representative specialized and general audio generators, selected according to task coverage, public availability, and recency. For the Complex-Scene evaluation, we retain recent speech-capable or unified systems with complementary task coverage: Higgs Audio V2, UniFlow-Audio, and Dasheng AudioGen. We exclude other baselines from this suite as they inherently lack either speech synthesis capabilities or general TTA support required for mixed scenes. 
We compare against representative specialized and general audio generators selected for task coverage, availability, and recency. For Complex-Scene evaluation, we retain Higgs Audio V2, UniFlow-Audio, and Dasheng AudioGen, as they provide the most relevant publicly available coverage of both speech generation and general audio generation for mixed scenes.

Two key ablation models, Dense (1B) and Base-MoE (2B, matching SonicWeave), share SonicWeave’s data and backbone. Dense replaces CPE-MoE with dense FFNs; Base-MoE uses token-level top-K audio routing without conflict-gated prior–evidence fusion. 
The Dense comparison measures the benefit of conditional capacity, whereas the parameter-matched Base-MoE comparison isolates the effect of the routing design more directly.

The complete model capability comparison is given in Table~\ref{tab:model_scope}, while task-specific quantitative comparisons are reported in Tables~\ref{tab:tts}--\ref{tab:ttm}.

\subsubsection{Metrics.}
For all public benchmarks, we report objective metrics suited to each task: WER/CER for TTS; FAD, KL divergence, and CLAP for TTA; and KL divergence and CLAP for TTM. The TTA and TTM KL/CLAP scores follow the Stable Audio metrics protocol with task-specific checkpoints, while TTA FAD uses the AVBench FD$_{\mathrm{VGG}}$ implementation. For the Complex-Scene evaluation, we use Gemini-based reference-free judging together with human listening tests. 
The two are used jointly because the reference-free judge provides scalable coverage over all prompts, while human MOS offers an independent perceptual assessment on a fixed subset.
% Gemini provides scalable coverage, while MOS calibrates absolute perceptual quality.
Gemini assigns 1-5 scores for semantic adherence and technical quality from the same prompt; the concise scoring rubric is given in the appendix.

\subsubsection{Implementation Details.}
% TODO: fill training config (params, layers, K CPE-MoE layers, N_e experts, top-K, chunk size C, optimizer, steps, data hours, hardware).
The CPE-MoE module replaces the dense FFN in the last $4$ layers, with $4$ routed experts and top-$2$ selection at chunk size $C{=}4$. Inference uses a fixed-step Euler ODE solver with APG guidance. Training uses roughly five million 10--15-second clips sampled from an internal corpus exceeding 20{,}000 hours, in which speech, music, sound effects, ambience, and mixed recordings coexist in a shared data distribution. Additional model and optimization details, together with targeted studies of gate adaptivity and chunk-size sensitivity, are provided in the appendix.

\subsection{Results on Public Benchmarks}
Tables~\ref{tab:tts}-\ref{tab:ttm} summarize the public-benchmark results. On TTS, SonicWeave achieves the best or tied-best result on all three subsets, tying the strongest reported results on SeedTTS-eval and LibriSpeech. Relative to the controlled baselines, it improves over Dense by 1.4, 1.0, and 2.5 percentage points on these subsets, and over Base-MoE by 0.4, 0.3, and 0.8 points, respectively.

\begin{table}[h]
\centering
\caption{TTS results on SeedTTS and LibriSpeech.}
\label{tab:tts}
\tabcolsep=1.8pt
\resizebox{\linewidth}{!}{%
\begin{tabular}{lccc}
\toprule
\multirow{2}{*}{Model} & \multicolumn{1}{c}{SeedTTS-en} & \multicolumn{1}{c}{SeedTTS-zh} & \multicolumn{1}{c}{LibriSpeech} \\
\cmidrule(lr){2-2}\cmidrule(lr){3-3}\cmidrule(lr){4-4}
 & WER$\downarrow$ & CER$\downarrow$ & WER$\downarrow$ \\
\midrule
F5-TTS       & 1.9\% & 1.6\% & \textbf{2.4\%} \\ % VERIFY
CosyVoice 2  & 2.0\% & 1.5\% & 2.5\% \\ % VERIFY
UniAudio     & 7.2\% & -- & 18.3\% \\ % VERIFY
Qwen2.5-Omni    & 2.1\% & 1.6\% & 7.6\% \\ % VERIFY
UniMoE-Audio & 1.3\% & \textbf{0.8\%} & 3.4\% \\ % VERIFY
Higgs Audio V2& \textbf{1.0\%} & \textbf{0.8\%} & 3.6\% \\ % VERIFY
Step-Audio & 2.2\% & 1.0\% & 5.0\% \\ % VERIFY
\midrule
Dense (ours)     & 2.4\% & 1.8\% & 4.9\% \\
Base-MoE (ours)   & 1.4\% & 1.1\% & 3.2\% \\
% \rowcolor{cyan!9}
SonicWeave (ours) & \textbf{1.0\%} & \textbf{0.8\%} & \textbf{2.4\%} \\
\bottomrule
\end{tabular}
}
\end{table}

On TTA, SonicWeave substantially improves over Dense and Base-MoE in all three metrics. Its CLAP score reaches 0.475, close to the strongest reported baseline (0.480), while its KL is 1.26 and its FAD is 2.75. 
% This indicates a balanced text--audio alignment and distributional match rather than only one metric.
SonicWeave does not obtain the best value on every individual AudioCaps metric, but it narrows the gap to the strongest CLAP system while substantially improving FAD and KL over the matched controls.

\begin{table}[h]
\centering
\caption{TTA results on AudioCaps.}
\label{tab:tta}
% \resizebox{\linewidth}{!}{%
\begin{tabular}{lccc}
\toprule
Model & FAD$\downarrow$ & KL$\downarrow$ & CLAP$\uparrow$ \\
\midrule
AudioLDM 2  & 2.29 & 1.41 & 0.419 \\ % VERIFY
TangoFlux   & \textbf{2.26} & \textbf{1.15} & \textbf{0.480} \\ % VERIFY
Stable Audio Open & 4.13 & 2.14 & 0.350 \\ % VERIFY
Dasheng AudioGen & 3.19 & 1.86 & 0.438 \\ % VERIFY
UniAudio          & 6.64 & -- & 0.243 \\ % VERIFY
UniFlow-Audio     & 5.74 & 1.43 & 0.476 \\ % VERIFY
AudioX            & 2.45 & 1.27 & 0.440 \\ % VERIFY
\midrule
Dense (ours)     & 3.98 & 1.62 & 0.389 \\
Base-MoE (ours)   & 3.10 & 1.40 & 0.422 \\
% \rowcolor{cyan!9}
SonicWeave (ours) & 2.75 & 1.26 & 0.475 \\
\bottomrule
\end{tabular}
% }
\end{table}

% On TTM, SonicWeave obtains the lowest KL on both MusicCaps and Song Describer, and the highest CLAP on Song Describer. On MusicCaps, its CLAP score of 0.312 is below UniMoE-Audio and Dasheng AudioGen but higher than the other listed baselines. Compared with Dense and Base-MoE, the consistent KL improvements and stronger CLAP results show that the proposed routing improves text--music correspondence across both datasets, without claiming uniform superiority over every specialized baseline.
On TTM, SonicWeave achieves the lowest KL and highest CLAP on Song Describer among the listed systems.
% On MusicCaps, it obtains a competitive KL of 1.06, while its CLAP score of 0.312 is below UniMoE-Audio and Dasheng AudioGen but higher than the other listed baselines. 
On MusicCaps, it achieves a competitive KL of 1.06 and CLAP of 0.312, offering a strong quality–alignment trade-off and placing it on the Pareto frontier among the listed baselines.
Compared with Dense and Base-MoE, SonicWeave consistently improves both KL and CLAP across the two datasets, showing that CPE-MoE strengthens text--music correspondence without uniformly surpassing every specialized baseline.

\begin{table}[h]
\centering
\caption{TTM results on MusicCaps and Song Describer.}
\label{tab:ttm}
\resizebox{\linewidth}{!}{%
\begin{tabular}{lcccc}
\toprule
\multirow{2}{*}{Model} & \multicolumn{2}{c}{MusicCaps} & \multicolumn{2}{c}{Song Describer} \\
\cmidrule(lr){2-3}\cmidrule(lr){4-5}
 & CLAP$\uparrow$ & KL$\downarrow$ & CLAP$\uparrow$ & KL$\downarrow$ \\
\midrule
AudioLDM 2  & 0.301 & 1.20 & -- & -- \\ % VERIFY
MusicGen-Large  & 0.280 & 1.31 & 0.19 & 0.54 \\ % VERIFY
Stable Audio Open   & 0.300 & 1.44 & 0.32 & 0.62 \\ % VERIFY
AudioX     & 0.240 & \textbf{0.96} & -- & -- \\ % VERIFY
Dasheng AudioGen     & 0.334 & 1.37 & -- & -- \\ % VERIFY
UniMoE-Audio     & \textbf{0.340} & 1.25 & 0.22 & -- \\ % VERIFY
UniFlow-Audio    & 0.241 & 1.87 & 0.15 & -- \\ % VERIFY
\midrule
Dense (ours)      & 0.251 & 1.31 & 0.29 & 0.58 \\
Base-MoE (ours)     & 0.296 & 1.10 & 0.32 & 0.48 \\
% \rowcolor{cyan!9}
SonicWeave (ours)   & 0.312 & 1.06 & \textbf{0.36} & \textbf{0.44} \\
\bottomrule
\end{tabular}
}
% }
\end{table}

\subsection{Complex-Scene Evaluation}
\begin{figure*}[t]
\centering
\includegraphics[width=0.54\linewidth]{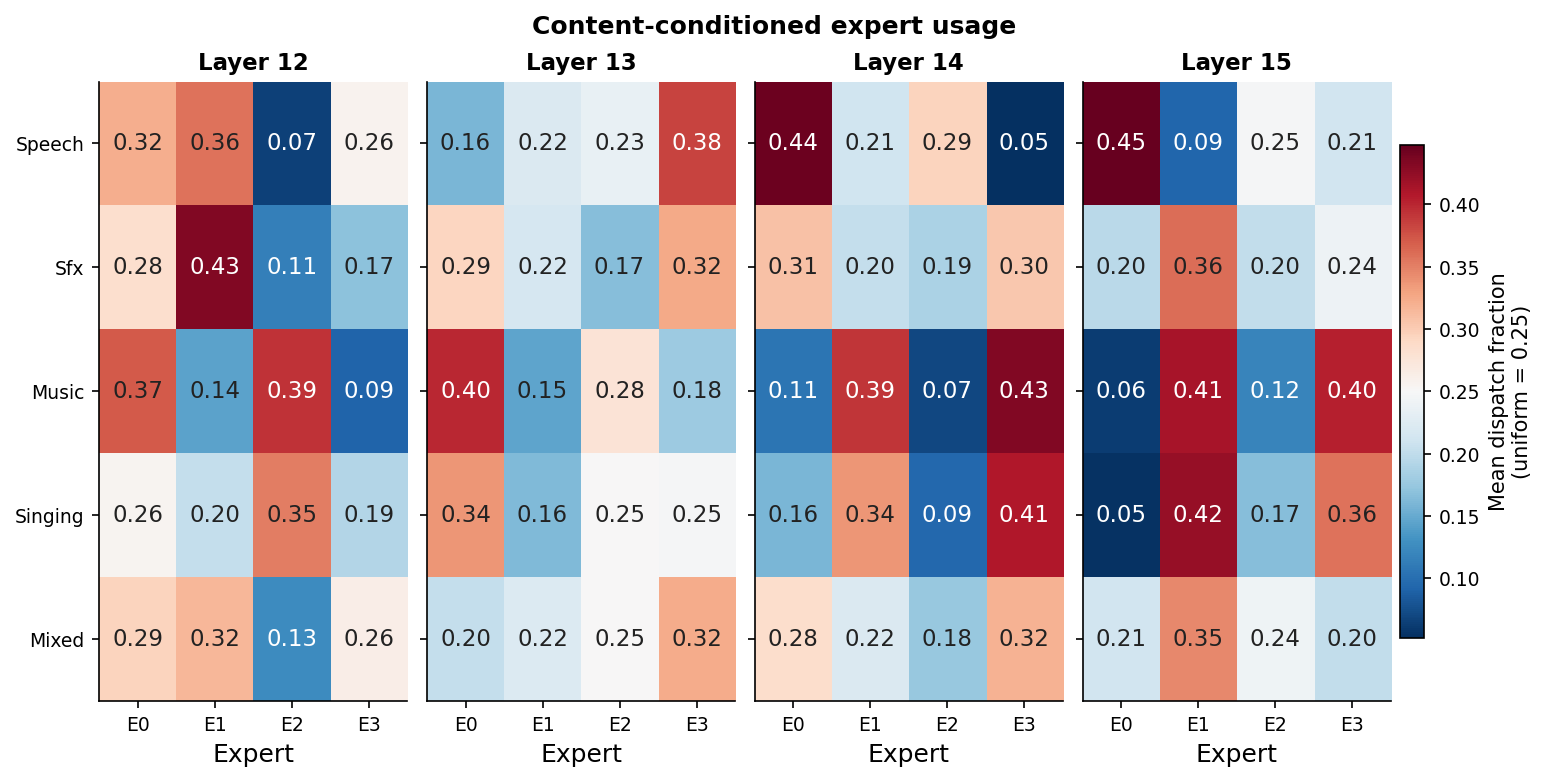}\hfill
\includegraphics[width=0.46\linewidth]{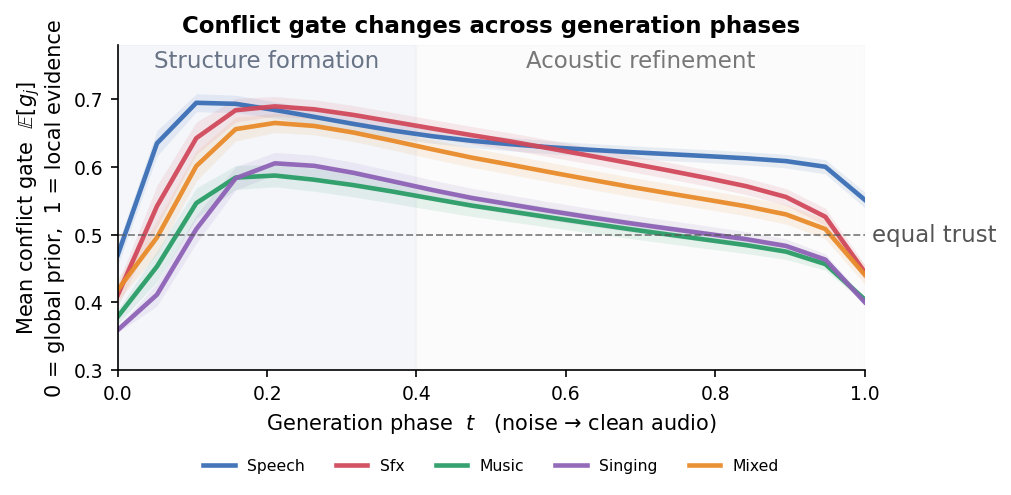}

% \vspace{-1mm}

\makebox[0.54\linewidth]{(a)}\hfill
\makebox[0.46\linewidth]{(b)}
\caption{Routing analysis on the balanced routing benchmark. (a) Content-conditioned expert dispatch across routed layers, showing non-uniform and content-dependent specialization. (b) Mean conflict gate $\mathbb{E}[g_j|t]$ across diffusion phases, showing content-dependent and phase-dependent prior--evidence balancing.}
\label{fig:routing}
\end{figure*}
Public benchmarks mainly evaluate isolated speech, music, or environmental sounds and therefore underrepresent the heterogeneous compositions targeted by SonicWeave. We construct a controlled suite of 100 manually curated complex-scene prompts, comprising 45 speech-with-background cases, 10 singing-with-background cases, 15 clean dialogues, and 30 dialogues with background. 
% We compare recent speech-capable or unified systems: Higgs Audio V2 is a recent speech-oriented reference, while UniFlow-Audio and Dasheng AudioGen provide unified-generation references. 
Each case is provided in both natural-language and structured-caption forms. 
% and specifies foreground content, vocal or speaker attributes, background components, and their acoustic relations. 
Public systems receive the natural-language prompt in their native input format, whereas SonicWeave, Dense and Base-MoE receive the corresponding structured caption containing the same
semantic specifications.
% We use Gemini-based reference-free judging and subjective MOS as evaluation metrics.
% ; Gemini assigns 1--5 scores for semantic adherence and technical quality using the same source prompt and scoring rubric for every system.

\begin{table}[h]
\centering
\caption{Reference-free and human listening evaluation on the Complex-Scene suite. AI-Tech and AI-Sem denote technical quality and semantic adherence, respectively; MOS-Q measures perceived audio quality, while MOS-R measures whether the requested prompt events are jointly realized. }
\label{tab:complex_scene}
\tabcolsep=1.0pt
% Values are reported as mean $\pm$ standard error across evaluation cases.}
\resizebox{\columnwidth}{!}{%
\begin{tabular}{lcccc}
\toprule
Model & AI-Tech$\uparrow$ & AI-Sem$\uparrow$
      & MOS-Q$\uparrow$ & MOS-R$\uparrow$ \\
\midrule
Higgs Audio V2        & 4.32 & 3.54  & 3.92$_{\pm 0.21}$ & 2.42$_{\pm 0.64}$ \\
UniFlow-Audio        & 2.28 & 1.77 & 1.99$_{\pm 0.33}$ & 1.87$_{\pm 0.80}$ \\
Dasheng AudioGen      & 3.10 & 3.07 & 2.97$_{\pm 0.41}$ & 3.25$_{\pm 0.73}$ \\
\midrule
Dense (ours)         & 4.13 & 4.23 & 4.29$_{\pm 0.24}$ & 4.15$_{\pm 0.37}$ \\
Base-MoE (ours)        & 4.62 & 4.58 & \textbf{4.57$_{\pm 0.13}$} & 4.31$_{\pm 0.31}$ \\
% \rowcolor{cyan!9}
SonicWeave (ours)    & \textbf{4.79}  & \textbf{4.72}
                     & \textbf{4.57$_{\pm 0.19}$} & \textbf{4.49$_{\pm 0.38}$} \\
\bottomrule
\end{tabular}
}
\end{table}

Table~\ref{tab:complex_scene} shows consistent trends between the reference-free AI evaluation and human listening tests. 
% The three matched in-house models substantially outperform the evaluated public systems, while 
SonicWeave achieves the highest AI semantic-adherence and technical-quality scores and the highest human MOS-R. Higgs Audio V2 retains relatively competitive perceived quality but receives a much lower request-realization score, showing that audio quality alone does not imply faithful realization of all requested scene components. 
% Dasheng AudioGen and UniFlow-Audio additionally exhibit larger rating variation across the heterogeneous languages and scene types. 
Dasheng AudioGen and UniFlow-Audio additionally show less consistent performance across the heterogeneous prompts.
SonicWeave and Base-MoE are effectively tied on MOS-Q, whereas SonicWeave is consistently stronger in semantic adherence and request realization. This supports the intended role of CPE-MoE: it preserves the acoustic quality of the underlying generator while improving compositional fidelity and fine-grained control in complex scenes.

% \subsection{Ablation Studies}
% We ablate the core design choices of the CPE-MoE module on [benchmark subset]: (i) routing granularity---token-level vs.\ chunk-level, and chunk size $C\!\in\!\{1,2,4,8\}$; (ii) routing signals---prior-only, evidence-only, and the full prior--evidence fusion; (iii) the conflict gate---replacing the learned gate $g_j$ with a fixed interpolation weight; (iv) diffusion-phase information in the prior---removing the time embedding from $\mathbf{p}$; and (v) the shared-expert / equal-weight fusion vs.\ a learnable convex weight. % TODO: fill ablation table once available.

% \begin{table}[t]
% \centering
% \caption{Ablation of CPE-MoE components. Metrics on [benchmark subset].}
% \label{tab:ablation}
% \begin{tabular}{lccc}
% \toprule
% Variant & [Metric 1] & [Metric 2] & [Metric 3] \\
% \midrule
% Token-level routing & -- & -- & -- \\
% Prior-only routing  & -- & -- & -- \\
% Evidence-only routing & -- & -- & -- \\
% Fixed gate ($g{=}0.5$) & -- & -- & -- \\
% w/o diffusion phase in prior & -- & -- & -- \\
% Learnable fusion weight & -- & -- & -- \\
% \textbf{Full SonicWeave} & -- & -- & -- \\
% \bottomrule
% \end{tabular}
% \end{table}

\subsection{Routing Analysis and Visualization}
\label{sec:routing}
To explain the behavior of SonicWeave beyond aggregate generation metrics, we analyze the routing patterns induced by its CPE-MoE module. The main-text visualization focuses on population-level evidence: content-conditioned expert usage and the evolution of the conflict gate across diffusion phases. Representative mixed-scene traces are reserved for the appendix because they illustrate the mechanism qualitatively rather than providing a statistically aggregated result.

For this analysis, we construct a 250-prompt routing benchmark by sampling examples from several single-task evaluation sets and combining them with complex-scene prompts. We group the resulting prompts into speech, music, singing, SFX, and mixed-scene categories, and aggregate routing statistics over all valid chunks and routed layers.

\subsubsection{Content-conditioned expert specialization.}
Figure~\ref{fig:routing}(a) shows the mean dispatch fraction for each expert across the four CPE-MoE layers and five content groups. The load is clearly non-uniform and content dependent: speech and SFX favor different expert combinations from music and singing, while mixed scenes retain a distinct distribution rather than collapsing to one of the single-content patterns. This specialization is also layer dependent. 
For example, music and singing place more mass on different experts in the deeper routed layers, whereas speech consistently emphasizes a smaller subset of experts.
These observations indicate that the MoE is using conditional capacity to form acoustic specializations, rather than merely distributing tokens uniformly across experts. 
% The figure is a routing-behavior diagnostic, not by itself a quality comparison; the latter is established by the matched Dense and BaseMoE results.

\subsubsection{Phase-dependent conflict gating.}
Figure~\ref{fig:routing}(b) reports $\mathbb{E}[g_j|t]$ for each content group. The curves are content dependent and non-monotonic: the gate changes rapidly during early structure formation, reaches content-specific operating ranges, and gradually returns toward a more balanced regime during late acoustic refinement. Speech and SFX maintain higher evidence reliance than music and singing over much of the trajectory, while mixed scenes lie between these regimes. Thus, the model does not apply a single fixed prior--evidence interpolation to all audio. Instead, it adjusts evidence reliance as a function of both acoustic content and diffusion phase, consistent with the intended role of the conflict gate.

% \subsubsection{Representative mixed-scene behavior.}
% A per-sample trace is useful for making the preceding population-level pattern concrete, but it should not be interpreted as a standalone performance result. In the appendix, we visualize a fixed mixed-scene example with the chunk-wise gate together with prior, evidence, and posterior expert assignments. A convincing example exhibits temporally coherent gate regions and posterior assignments that remain close to the prior when the gate is low but follow local evidence when the gate increases. This case-level view complements, rather than replaces, the aggregate statistics in Figure~\ref{fig:routing}; it also avoids using one hand-selected sample as evidence for general generation quality.

\section{Conclusion}
This paper presented SonicWeave, a unified flow-matching model for generating speech, singing, music, sound effects, and their mixtures with a single set of parameters. The central premise of this work is that conditional computation in unified audio must operate at two coupled scales: it must accommodate differences across acoustic domains while adapting to heterogeneous regions within the same scene. CPE-MoE addresses this requirement by routing temporally contiguous audio chunks through a learned combination of a global text-and-diffusion prior and local acoustic evidence, while preserving a shared conditioning pathway for text and time tokens.

Across public TTS, text-to-audio, and text-to-music benchmarks, SonicWeave consistently improves over matched Dense and token-routed MoE controls. On the Complex-Scene suite, the principal gain is compositional: SonicWeave improves semantic adherence and joint request realization while maintaining perceptual quality comparable to the parameter-matched Base-MoE. 
Quantitatively, it reduces TTA FAD by 30.9\% over Dense and improves Complex-Scene MOS-R by 0.18 over Base-MoE and by more than 1.2 points over the strongest external baseline.
The learned expert assignments and conflict-gate trajectories also vary with content, layer depth, and diffusion phase, providing evidence that the additional capacity is used conditionally rather than uniformly.

Targeted studies reported in the appendix further compare the learned gate with fixed prior–evidence averaging and vary the routing chunk size. Fixed fusion degrades all reported task groups, while increasing the chunk size from 4 to 8 weakens most metrics, supporting adaptive fusion and relatively fine-grained chunk routing. 
% These controls are targeted rather than exhaustive: they do not separately isolate the prior-only and evidence-only branches.

% Across TTS, TTA, and TTM benchmarks, SonicWeave consistently outperforms the controlled Dense and Base-MoE baselines. It achieves best or tied-best performance across all three TTS subsets, obtains the lowest KL and highest CLAP on Song Describer among the listed systems, and surpasses strong unified audio generators on the Complex-Scene benchmark.
% Quantitatively, it reduces TTA FAD by 30.9\% over Dense and improves Complex-Scene MOS-R by 0.18 over Base-MoE and by more than 1.2 points over the strongest external baseline.

These findings support chunk-level prior–evidence routing as a useful intermediate granularity for unified audio generation. At the same time, the current evaluation is limited to short-form generation, a two-speaker structured interface, and non-exhaustive routing ablations. Extending the approach to longer scenes, richer speaker interactions, and explicit temporal plans remains an important direction.
% This paper presented SonicWeave, a unified flow-matching model whose core CPE-MoE module routes contiguous acoustic chunks by fusing a text-and-phase prior with local acoustic evidence through a learned conflict gate. 

% % Across TTS, TTA, and TTM benchmarks, SonicWeave consistently outperforms matched Dense and Base-MoE controls, achieves the lowest KL divergence on both MusicCaps and Song Describer among all reported systems, and delivers the highest human MOS on the Complex-Scene suite, surpassing strong external unified baselines such as Higgs Audio V2 and Dasheng AudioGen. 
% Across TTS, TTA, and TTM benchmarks, SonicWeave consistently outperforms the controlled Dense and Base-MoE baselines. It achieves best or tied-best performance across all three TTS subsets, obtains the lowest KL and highest CLAP on Song Describer among the listed systems, and surpasses strong unified audio generators on the Complex-Scene benchmark.
% Quantitatively, it reduces TTA FAD by 30.9\% over Dense and improves Complex-Scene MOS-R by 0.18 over Base-MoE and by more than 1.2 points over the strongest external baseline.
% These results support the role of CPE-MoE in preserving unified generation while enabling fine-grained specialization.

% Future work will extend the structured caption interface to richer multi-speaker and more diverse compositional scenes, while further decomposing the routing signals for longer-form generation.

\bibliographystyle{unsrtnat}
\bibliography{aaai2027}

@article{voicebox,
  title={Voicebox: Text-guided multilingual universal speech generation at scale},
  author={Le, Matthew and Vyas, Apoorv and Shi, Bowen and Karrer, Brian and Sari, Leda and Moritz, Rashel and Williamson, Mary and Manohar, Vimal and Adi, Yossi and Mahadeokar, Jay and others},
  journal={Advances in neural information processing systems},
  volume={36},
  pages={14005--14034},
  year={2023}
}

@inproceedings{f5tts,
  author    = {Yushen Chen and Zhikang Niu and Ziyang Ma and Keqi Deng and Chunhui Wang and Jian Zhao and Kai Yu and Xie Chen},
  title     = {{F5-TTS}: A Fairytaler that Fakes Fluent and Faithful Speech with Flow Matching},
  booktitle = {Proceedings of the 63rd Annual Meeting of the Association for Computational Linguistics (Volume 1: Long Papers)},
  pages     = {6255--6271},
  year      = {2025}
}

@inproceedings{tangoflux,
title={TangoFlux: Super Fast and Faithful Text to Audio Generation with Flow Matching and Clap-Ranked Preference Optimization},
author={Chia-Yu Hung and Navonil Majumder and Zhifeng Kong and Ambuj Mehrish and Amir Zadeh and Chuan Li and Rafael Valle and Bryan Catanzaro and Soujanya Poria},
booktitle={The Fourteenth International Conference on Learning Representations},
year={2026},
pages     = {1--24}
}

@article{audioldm2,
  author  = {Haohe Liu and Yi Yuan and Xubo Liu and Xinhao Mei and Qiuqiang Kong and Qiao Tian and Yuping Wang and Wenwu Wang and Yuxuan Wang and Mark D. Plumbley},
  title   = {{AudioLDM 2}: Learning Holistic Audio Generation With Self-Supervised Pretraining},
  journal = {IEEE/ACM Transactions on Audio, Speech, and Language Processing},
  volume  = {32},
  pages   = {2871--2883},
  year    = {2024}
}

@inproceedings{uniaudio,
  author    = {Dongchao Yang and Jinchuan Tian and Xu Tan and Rongjie Huang and Songxiang Liu and Haohan Guo and Xuankai Chang and Jiatong Shi and Sheng Zhao and Jiang Bian and Zhou Zhao and Xixin Wu and Helen M. Meng},
  title     = {{UniAudio}: Towards Universal Audio Generation with Large Language Models},
  booktitle = {Proceedings of the 41st International Conference on Machine Learning},
  pages     = {56422--56447},
  year      = {2024}
}

@article{audiobox,
  author  = {Apoorv Vyas and Bowen Shi and Matthew Le and Andros Tjandra and Yi{-}Chiao Wu and Baishan Guo and Jiemin Zhang and Xinyue Zhang and Robert Adkins and William Ngan and Jeff Wang and Ivan Cruz and Bapi Akula and Akinniyi Akinyemi and Brian Ellis and Rashel Moritz and Yael Yungster and Alice Rakotoarison and Liang Tan and Chris Summers and Carleigh Wood and Joshua Lane and Mary Williamson and Wei{-}Ning Hsu},
  title   = {{Audiobox}: Unified Audio Generation with Natural Language Prompts},
  journal = {CoRR},
  volume  = {abs/2312.15821},
  year    = {2023},
  doi     = {10.48550/arXiv.2312.15821}
}

@inproceedings{audioldm,
  author    = {Haohe Liu and Zehua Chen and Yi Yuan and Xinhao Mei and Xubo Liu and Danilo P. Mandic and Wenwu Wang and Mark D. Plumbley},
  title     = {{AudioLDM}: Text-to-Audio Generation with Latent Diffusion Models},
  booktitle = {Proceedings of the 40th International Conference on Machine Learning},
  pages     = {21450--21474},
  year      = {2023}
}

@inproceedings{stableaudio,
  author    = {Zach Evans and Julian D. Parker and CJ Carr and Zack Zukowski and Josiah Taylor and Jordi Pons},
  title     = {Stable Audio Open},
  booktitle = {IEEE International Conference on Acoustics, Speech and Signal Processing},
  pages     = {1--5},
  year      = {2025}
}

@article{cosyvoice2,
  author  = {Zhihao Du and Yuxuan Wang and Qian Chen and Xian Shi and Xiang Lv and Tianyu Zhao and Zhifu Gao and Yexin Yang and Changfeng Gao and Hui Wang and Fan Yu and Huadai Liu and Zhengyan Sheng and Yue Gu and Chong Deng and Wen Wang and Shiliang Zhang and Zhijie Yan and Jingren Zhou},
  title   = {{CosyVoice 2}: Scalable Streaming Speech Synthesis with Large Language Models},
  journal = {CoRR},
  volume  = {abs/2412.10117},
  year    = {2024},
  doi     = {10.48550/arXiv.2412.10117}
}

@article{makeanaudio2,
  author  = {Jiawei Huang and Yi Ren and Rongjie Huang and Dongchao Yang and Zhenhui Ye and Chen Zhang and Jinglin Liu and Xiang Yin and Zejun Ma and Zhou Zhao},
  title   = {{Make-An-Audio 2}: Temporal-Enhanced Text-to-Audio Generation},
  journal = {CoRR},
  volume  = {abs/2305.18474},
  year    = {2023},
  doi     = {10.48550/arXiv.2305.18474}
}

@inproceedings{tango,
  title={Text-to-audio generation using instruction guided latent diffusion model},
  author={Ghosal, Deepanway and Majumder, Navonil and Mehrish, Ambuj and Poria, Soujanya},
  booktitle={Proceedings of the 31st ACM international conference on multimedia},
  pages={3590--3598},
  year={2023}
}

@inproceedings{freeaudio,
  title={Freeaudio: Training-free timing planning for controllable long-form text-to-audio generation},
  author={Jiang, Yuxuan and Chen, Zehua and Ju, Zeqian and Li, Chang and Dou, Weibei and Zhu, Jun},
  booktitle={Proceedings of the 33rd ACM International Conference on Multimedia},
  pages={9871--9880},
  year={2025}
}

@inproceedings{unisonate,
  author    = {Chunyu Qiang and Xiaopeng Wang and Kang Yin and Yuzhe Liang and Yuxin Guo and Teng Ma and Ziyu Zhang and Tianrui Wang and Cheng Gong and Yushen Chen and Ruibo Fu and Longbiao Wang and Jianwu Dang},
  title     = {{UniSonate}: A Unified Model for Speech, Music, and Sound Effect Generation with Text Instructions},
  booktitle = {Proceedings of the 64th Annual Meeting of the Association for Computational Linguistics},
  pages     = {28043--28054},
  year      = {2026}
}

@article{dashengaudiogen,
  author  = {Jiahao Mei and Heinrich Dinkel and Yadong Niu and Xingwei Sun and Gang Li and Yifan Liao and Jiahao Zhou and Junbo Zhang and Jian Luan and Mengyue Wu},
  title   = {{Dasheng AudioGen}: A Unified Model for Generating Coherent Audio Scenes from Text},
  journal = {CoRR},
  volume  = {abs/2605.27838},
  year    = {2026},
  doi     = {10.48550/arXiv.2605.27838}
}

@article{switchtransformer,
  author  = {William Fedus and Barret Zoph and Noam Shazeer},
  title   = {Switch Transformers: Scaling to Trillion Parameter Models
             with Simple and Efficient Sparsity},
  journal = {Journal of Machine Learning Research},
  volume  = {23},
  number  = {120},
  pages   = {1--39},
  year    = {2022}
}

@article{ditmoe,
  author  = {Zhengcong Fei and Mingyuan Fan and Changqian Yu and Debang Li and Junshi Huang},
  title   = {Scaling Diffusion Transformers to 16 Billion Parameters},
  journal = {CoRR},
  volume  = {abs/2407.11633},
  year    = {2024},
  doi     = {10.48550/arXiv.2407.11633}
}

@inproceedings{switchdit,
  title={Switch diffusion transformer: Synergizing denoising tasks with sparse mixture-of-experts},
  author={Park, Byeongjun and Go, Hyojun and Kim, Jin-Young and Woo, Sangmin and Ham, Seokil and Kim, Changick},
  booktitle={European Conference on Computer Vision},
  pages={461--477},
  year={2024},
  organization={Springer}
}

@inproceedings{ecdit,
  author    = {Haotian Sun and Tao Lei and Bowen Zhang and Yanghao Li and Haoshuo Huang and Ruoming Pang and Bo Dai and Nan Du},
  title     = {{EC-DIT}: Scaling Diffusion Transformers with Adaptive Expert-Choice Routing},
  booktitle = {The Thirteenth International Conference on Learning Representations},
  year      = {2025}
}

@inproceedings{unimoeaudio,
  author    = {Zhenyu Liu and Yunxin Li and Xuanyu Zhang and Qixun Teng and Shenyuan Jiang and Xinyu Chen and Haoyuan Shi and Haolan Chen and Fanbo Meng and Mingjun Zhao and Yu Xu and Yancheng He and Baotian Hu and Haizhou Li and Min Zhang},
  title     = {{UniMoE-Audio}: Unified Speech and Music Generation with Dynamic-Capacity Mixture-of-Experts},
  booktitle = {Proceedings of the 64th Annual Meeting of the Association for Computational Linguistics (Volume 1: Long Papers)},
  pages     = {9107--9119},
  year      = {2026}
}

@article{seedtts,
  author  = {Philip Anastassiou and Jiawei Chen and Jitong Chen and Yuanzhe Chen and Zhuo Chen and Ziyi Chen and Jian Cong and Lelai Deng and Chuang Ding and Lu Gao and Mingqing Gong and Peisong Huang and Qingqing Huang and Zhiying Huang and Yuanyuan Huo and Dongya Jia and Chumin Li and Feiya Li and Hui Li and Jiaxin Li and Xiaoyang Li and Xingxing Li and Lin Liu and Shouda Liu and Sichao Liu and Xudong Liu and Yuchen Liu and Zhengxi Liu and Lu Lu and Junjie Pan and Xin Wang and Yuping Wang and Yuxuan Wang and Zhen Wei and Jian Wu and Chao Yao and Yifeng Yang and Yuanhao Yi and Junteng Zhang and Qidi Zhang and Shuo Zhang and Wenjie Zhang and Yang Zhang and Zilin Zhao and Dejian Zhong and Xiaobin Zhuang},
  title   = {{Seed-TTS}: A Family of High-Quality Versatile Speech Generation Models},
  journal = {CoRR},
  volume  = {abs/2406.02430},
  year    = {2024},
  doi     = {10.48550/arXiv.2406.02430}
}

@inproceedings{apg,
  title={Eliminating oversaturation and artifacts of high guidance scales in diffusion models},
  author={Sadat, Seyedmorteza and Hilliges, Otmar and Weber, Romann M},
  booktitle={The Thirteenth International Conference on Learning Representations},
  year={2025}
}

@article{cfg,
  title={Classifier-free diffusion guidance},
  author={Ho, Jonathan and Salimans, Tim},
  journal={arXiv preprint arXiv:2207.12598},
  year={2022}
}

@inproceedings{
flow,
title={Flow Matching for Generative Modeling},
author={Yaron Lipman and Ricky T. Q. Chen and Heli Ben-Hamu and Maximilian Nickel and Matthew Le},
booktitle={The Eleventh International Conference on Learning Representations },
year={2023},
pages = {1--28}
}

@inproceedings{librispeechpc,
  title     = {LibriSpeech-PC: Benchmark for Evaluation of
               Punctuation and Capitalization Capabilities of
               End-to-End ASR Models},
  author    = {Aleksandr Meister and Matvei Novikov and Nikolay Karpov
               and Evelina Bakhturina and Vitaly Lavrukhin
               and Boris Ginsburg},
  booktitle = {2023 IEEE Automatic Speech Recognition and
               Understanding Workshop (ASRU)},
  pages     = {1--7},
  year      = {2023}
}

@inproceedings{audiocaps,
  author    = {Chris Dongjoo Kim and Byeongchang Kim and Hyunmin Lee and Gunhee Kim},
  title     = {{AudioCaps}: Generating Captions for Audios in the Wild},
  booktitle = {Proceedings of the 2019 Conference of the North American Chapter of the Association for Computational Linguistics: Human Language Technologies, Volume 1},
  pages     = {119--132},
  year      = {2019}
}

@article{musiccaps,
  author  = {Andrea Agostinelli and Timo I. Denk and Zal{\'a}n Borsos and Jesse H. Engel and Mauro Verzetti and Antoine Caillon and Qingqing Huang and Aren Jansen and Adam Roberts and Marco Tagliasacchi and Matthew Sharifi and Neil Zeghidour and Christian Havn{\o} Frank},
  title   = {{MusicLM}: Generating Music From Text},
  journal = {CoRR},
  volume  = {abs/2301.11325},
  year    = {2023},
  doi     = {10.48550/arXiv.2301.11325}
}

@article{songdescriber,
  author  = {Ilaria Manco and Benno Weck and Seungheon Doh and Minz Won and Yixiao Zhang and Dmitry Bogdanov and Yusong Wu and Ke Chen and Philip Tovstogan and Emmanouil Benetos and Elio Quinton and Gy{\"o}rgy Fazekas and Juhan Nam},
  title   = {The Song Describer Dataset: A Corpus of Audio Captions for Music-and-Language Evaluation},
  journal = {CoRR},
  volume  = {abs/2311.10057},
  year    = {2023},
  doi     = {10.48550/arXiv.2311.10057}
}

@article{musicgen,
  title={Simple and controllable music generation},
  author={Copet, Jade and Kreuk, Felix and Gat, Itai and Remez, Tal and Kant, David and Synnaeve, Gabriel and Adi, Yossi and D{\'e}fossez, Alexandre},
  journal={Advances in neural information processing systems},
  volume={36},
  pages={47704--47720},
  year={2023}
}

@article{uniflowaudio,
  author  = {Xuenan Xu and Jiahao Mei and Zihao Zheng and Ye Tao and Zeyu Xie and Yaoyun Zhang and Haohe Liu and Yuning Wu and Ming Yan and Wen Wu and Chao Zhang and Mengyue Wu},
  title   = {{UniFlow-Audio}: Unified Flow Matching for Audio Generation from Omni-Modalities},
  journal = {CoRR},
  volume  = {abs/2509.24391},
  year    = {2025},
  doi     = {10.48550/arXiv.2509.24391}
}

@inproceedings{infersent,
  author    = {Alexis Conneau and Douwe Kiela and Holger Schwenk and Lo{\"i}c Barrault and Antoine Bordes},
  title     = {Supervised Learning of Universal Sentence Representations from Natural Language Inference Data},
  booktitle = {Proceedings of the 2017 Conference on Empirical Methods in Natural Language Processing},
  pages     = {670--680},
  publisher = {Association for Computational Linguistics},
  year      = {2017},
  doi       = {10.18653/v1/D17-1070}
}

@article{qwen2vl,
  author  = {Peng Wang and Shuai Bai and Sinan Tan and Shijie Wang and Zhihao Fan and Jinze Bai and Keqin Chen and Xuejing Liu and Jialin Wang and Wenbin Ge and Yang Fan and Kai Dang and Mengfei Du and Xuancheng Ren and Rui Men and Dayiheng Liu and Chang Zhou and Jingren Zhou and Junyang Lin},
  title   = {{Qwen2-VL}: Enhancing Vision-Language Model's Perception of the World at Any Resolution},
  journal = {CoRR},
  volume  = {abs/2409.12191},
  year    = {2024},
  doi     = {10.48550/arXiv.2409.12191}
}

@article{qwen25omni,
  author  = {Jin Xu and Zhifang Guo and Jinzheng He and Hangrui Hu and Ting He and Shuai Bai and Keqin Chen and Jialin Wang and Yang Fan and Kai Dang and Bin Zhang and Xiong Wang and Yunfei Chu and Junyang Lin},
  title   = {{Qwen2.5-Omni} Technical Report},
  journal = {CoRR},
  volume  = {abs/2503.20215},
  year    = {2025},
  doi     = {10.48550/arXiv.2503.20215}
}

@inproceedings{audiox,
title={AudioX: A Unified Framework for Anything-to-Audio Generation},
author={Zeyue Tian and Zhaoyang Liu and Yizhu Jin and Ruibin Yuan and Liumeng Xue and Xu Tan and Qifeng Chen and Wei Xue and Yike Guo},
booktitle={The Fourteenth International Conference on Learning Representations},
year={2026},
pages     = {1--27}
}

@misc{higgsaudio,
  author       = {{Boson AI}},
  title        = {{Higgs Audio V2}: Redefining Expressiveness in Audio Generation},
  year         = {2025},
  howpublished = {GitHub repository},
  note         = {Boson AI Higgs Audio project}
}

@inproceedings{segtune,
  author    = {Yuejiao Wang and Zihao Ji and Pengfei Cai and Xu Li and
               Haorui Zheng and Zewen Song and Zhongliang Liu and
               Chen Zhang and Pengfei Wan},
  title     = {{SegTune}: Structured and Fine-Grained Control for Song Generation},
  booktitle = {Proceedings of the 64th Annual Meeting of the Association for Computational Linguistics (Volume 1: Long Papers)},
  pages     = {12883--12897},
  year      = {2026},
  publisher = {Association for Computational Linguistics},
  doi       = {10.18653/v1/2026.acl-long.586}
}

@article{stepaudio,
  author  = {Ailin Huang and Boyong Wu and Bruce Wang and Chao Yan and
             Chen Hu and Chengli Feng and Fei Tian and Feiyu Shen and
             Jingbei Li and Mingrui Chen and others},
  title   = {{Step-Audio}: Unified Understanding and Generation in
             Intelligent Speech Interaction},
  journal = {CoRR},
  volume  = {abs/2502.11946},
  year    = {2025},
  doi     = {10.48550/arXiv.2502.11946}
}

@inproceedings{dit,
  author    = {William Peebles and Saining Xie},
  title     = {Scalable Diffusion Models with Transformers},
  booktitle = {Proceedings of the IEEE/CVF International Conference on Computer Vision},
  pages     = {4195--4205},
  year      = {2023}
}

@inproceedings{clip,
  author    = {Alec Radford and Jong Wook Kim and Chris Hallacy and Aditya Ramesh and Gabriel Goh and Sandhini Agarwal and Girish Sastry and Amanda Askell and Pamela Mishkin and Jack Clark and Gretchen Krueger and Ilya Sutskever},
  title     = {Learning Transferable Visual Models From Natural Language Supervision},
  booktitle = {Proceedings of the 38th International Conference on Machine Learning},
  pages     = {8748--8763},
  year      = {2021}
}

@article{qwen3-vl,
  title={Qwen3-vl technical report},
  author={Bai, Shuai and Cai, Yuxuan and Chen, Ruizhe and Chen, Keqin and Chen, Xionghui and Cheng, Zesen and Deng, Lianghao and Ding, Wei and Gao, Chang and Ge, Chunjiang and others},
  journal={arXiv preprint arXiv:2511.21631},
  year={2025}
}

@inproceedings{whisper,
  author    = {Alec Radford and Jong Wook Kim and Tao Xu and Greg Brockman and Christine McLeavey and Ilya Sutskever},
  title     = {Robust Speech Recognition via Large-Scale Weak Supervision},
  booktitle = {Proceedings of the 40th International Conference on Machine Learning},
  series    = {Proceedings of Machine Learning Research},
  volume    = {202},
  pages     = {28492--28518},
  year      = {2023}
}

@inproceedings{paraformer,
  author    = {Zhifu Gao and Shiliang Zhang and Ian McLoughlin and Zhijie Yan},
  title     = {Paraformer: Fast and Accurate Parallel Transformer for Non-Autoregressive End-to-End Speech Recognition},
  booktitle = {Interspeech 2022},
  pages     = {2063--2067},
  year      = {2022},
  doi       = {10.21437/Interspeech.2022-9996}
}

\clearpage
\appendix
\section{Architecture and Objective Details}
\label{app:architecture}

\subsection{Phase-Aware Text Pooling and AdaLN Conditioning}

Let $\mathbf{E}\in\mathbb{R}^{B\times L\times d_E}$ be the frozen text-encoder states, where $d_E=4096$, and let $\mathbf{m}\in\{0,1\}^{B\times L}$ be the caption-validity mask.  The diffusion-time and segment-start embeddings are denoted by $\mathbf{t}$ and $\mathbf{s}$, respectively.  We first construct the phase condition
\begin{equation}
\mathbf{u}=\mathbf{t}+\mathbf{s}\in\mathbb{R}^{B\times d_c},\qquad d_c=512.
\end{equation}
Rather than mean-pooling all text tokens, SonicWeave uses $\mathbf{u}$ as a query that changes with the generation phase:
\begin{equation}
\begin{aligned}
\mathbf{q}&=W_q\mathbf{u}, & \mathbf{K}&=W_k\mathbf{E},\\
a_i&=\frac{\mathbf{q}^{\top}\mathbf{K}_i}{\sqrt{d_E}}, &
\alpha_i&=\frac{\exp(a_i)\,m_i}{\sum_{r=1}^{L}\exp(a_r)\,m_r+\epsilon},\\
\bar{\mathbf{e}}&=\sum_{i=1}^{L}\alpha_i\mathbf{E}_i, &
\mathbf{c}&=\mathbf{u}+W_g\bar{\mathbf{e}}.
\end{aligned}
\label{eq:adaln-pooling}
\end{equation}
The implementation masks invalid text positions before the softmax, multiplies the resulting weights by $\mathbf{m}$ once more, and renormalizes.  Consequently, padding contributes neither to $\bar{\mathbf{e}}$ nor to its gradients.  If all text tokens are absent (the caption-dropped classifier-free guidance branch), $\bar{\mathbf{e}}$ is set to zero and $\mathbf{c}=\mathbf{u}$.

The global condition $\mathbf{c}$ is supplied to the final AdaLN-Zero layer that normalizes the audio-token states before the velocity projection.  Token-level text--audio correspondence is still modeled through joint self-attention; Eq.~\eqref{eq:adaln-pooling} is a complementary global path.  Its phase-dependent query lets the model alter which parts of a structured caption are globally emphasized during generation, while the residual $\mathbf{u}$ preserves explicit diffusion and segment-position information.

\subsection{Backbone and CPE-MoE Configuration}

Table~\ref{tab:architecture-config} lists the configuration used by the reported SonicWeave model.  The model consumes a sequence consisting of one time token, up to 200 text tokens in the reported configuration, and at most 256 audio latent frames.  Text and audio positions use independent coordinate ranges under rotary positional embedding, so an audio position does not change when the caption length changes.  The audio input projection receives the concatenation of the current noised latent and the masked reference latent; the velocity head predicts 64 channels per latent frame.

\begin{table}[h]
\centering
\caption{Backbone and CPE-MoE configuration.  The text encoder is frozen; all listed DiT and routing components are trained jointly.}
\label{tab:architecture-config}
\begin{tabular}{@{}p{0.36\linewidth}p{0.56\linewidth}@{}}
\toprule
Component & Configuration \\
\midrule
Text encoder & Frozen Qwen3-VL-8B~\cite{qwen3-vl} text encoder; $d_E=4096$ \\
DiT width / depth & $d=2048$ / 16 transformer blocks \\
Attention & 32 heads; head width 64; SDPA attention \\
Dense FFN width & $4d=8192$; SiLU activation \\
Text-token limit & 200 tokens \\
Max audio length & 256 latent frames \\
Latent channels & 64 channels \\
AdaLN width & 512 \\
Routed blocks & Final 4 of 16 blocks \\
Routed experts & 4 experts per routed block \\
Sparse selection & Top-2 experts per valid chunk \\
Chunk length & $C=4$ latent frames \\
CL Loss & $\alpha_{\mathrm{CL}}$=0.1 \\
Load Balancing & $\alpha_{\mathrm{MoE}}$=0.05 \\
\bottomrule
\end{tabular}
\end{table}

Each routed block retains one shared FFN.  The time token and all text tokens are processed only by this shared FFN.  Only audio tokens enter the routed path, which prevents the routing decision from directly modifying the condition-token pathway.  For a valid chunk, top-2 posterior probabilities are renormalized to sum to one, broadcast to all frames in the chunk, and used to weight the corresponding expert outputs.  Expert outputs are computed on the original frame states, not on pooled chunk states; pooling is used only to make a routing decision.

The routed experts use zero-initialized output projections.  Thus, the routed contribution begins near zero and is introduced smoothly as optimization learns both the experts and the router.  For audio tokens, the shared and sparse paths are combined with equal weights, $\tfrac12\mathbf{o}_{\mathrm{shared}}+\tfrac12\mathbf{o}_{\mathrm{sparse}}$.  Padding frames and padding-only chunks are masked before dispatch and are excluded from all router statistics.

\section{Structured Prompt Interface}
\label{app:structured_prompts}

\subsection{Schema and Field Semantics}

SonicWeave uses one structured text-conditioning interface for all text-to-audio generation modes.  A prompt consists of a short natural-language summary followed by a sparse set of XML-like fields (shown in Table~\ref{tab:prompt-schema}).  Fields with no applicable information are omitted rather than filled with generic text.  This is important: absence is meaningful for, for example, instrumental music, which should not receive an invented transcript or vocal attribute.

\begin{table*}[t]
\centering
\caption{Structured prompt fields.  The interface describes requested audible attributes only.}
\label{tab:prompt-schema}
\begin{tabular}{p{0.16\linewidth}p{0.25\linewidth}p{0.51\linewidth}}
\toprule
Field & Value type & Semantics \\
\midrule
\texttt{type} & categorical & Dominant audio mode: speech, singing, music, sound effects, or mixed.  \texttt{mixed} is used only when two or more of speech, singing, music, and sound effects are explicitly requested; it does not itself imply speech. \\
\texttt{lang} & language code & Language of intelligible speech or singing.  It is omitted for non-vocal audio. \\
\texttt{speech} & verbatim text & Spoken linguistic content, including ordered dialogue turns separated by slashes when applicable. \\
\texttt{lyrics} & text & Song lyrics content when provided.  \\
\texttt{speaker\_count} & categorical & Number of foreground speakers or vocalists when applicable: 1 or 2.  Dialogue scenes support at most two speakers. \\
\texttt{relation} & short phrase & Speaker roles, turn-taking relation, or foreground interaction, such as a host and guest or two commuters. \\
\texttt{vocal} & short phrase & Speaker identity descriptors, vocal timbre, delivery, emotion, accent, or recording coloration relevant to the foreground voice. \\
\texttt{music} & short phrase & Instrumentation, genre, tempo, mood, and relative role of music in the scene. \\
\texttt{sfx} & short phrase & Discrete non-musical events, such as a door slam, train brakes, keyboard taps, or rain impacts. \\
\texttt{ambience} & short phrase & Persistent environment and room-scale context, such as an indoor station, kitchen room tone, or open coastline. \\
\texttt{texture} & short phrase & Mix and recording relationships: foreground prominence, masking, reverberation, stereo impression, or dynamic behavior. \\
\bottomrule
\end{tabular}
\end{table*}

The division between \texttt{sfx}, \texttt{ambience}, and \texttt{texture} is deliberate.  \texttt{sfx} identifies events that must be audibly realized, \texttt{ambience} identifies the persistent environment, and \texttt{texture} specifies how components should coexist.  For instance, ``rain'' may be an event in \texttt{sfx}, ``an urban street in a downpour'' belongs in \texttt{ambience}, and ``rain intermittently masks dialogue'' belongs in \texttt{texture}.  This factorization makes foreground content, background content, and their acoustic relationship separately addressable by the text encoder and the CPE-MoE prior.

\subsection{Natural-to-Structured Conversion}
\label{app:nl_to_structured}

SonicWeave uses a unified structured audio caption interface to represent heterogeneous acoustic content. The goal of this conversion is not to rewrite or enrich natural-language descriptions, but to normalize diverse descriptions into a fixed set of fields that can be processed consistently by the text conditioning module. The structured representation covers acoustic type, language, speech content or lyrics, speaker-related attributes, music, sound events, ambience, and recording texture. Thus, “rule-based templates” refers to the fixed conversion rules and output schema, rather than unconstrained caption rewriting.

We formulate the conversion as constrained information extraction. Given a natural-language annotation, the converter retains the core summary supplied by the source annotation in the \texttt{summary} field and maps the remaining explicitly stated information into the predefined structured fields. Quoted speech and lyrics are copied verbatim when available, while unsupported fields remain empty. The converter is prohibited from introducing new events, speaker identities, emotions, acoustic properties, or stylistic attributes that are not stated in the source description.

The conversion follows a fixed schema and deterministic field constraints. When automatic conversion is required, we implement the constrained extractor with a language model (Qwen3.5-35B-A3B with temperature set to $0.0$) as a schema-aware mapper and validator: it applies the predefined rules, formats the output into the required structure, and checks for invalid or unsupported fields. It does not perform free-form caption generation. This design allows both training annotations and inference prompts to share the same structured interface while preserving the semantics of the original natural-language input.

The complete conversion instruction used by the extractor is provided
below.

\begin{promptbox}[title={Natural-to-Structured Conversion Prompt}]
\small % 如果内容太长，可以在这里统一将字体缩小
You are a structured audio metadata extractor for a text-to-audio
model.

You receive a natural-language description of an audio clip. Your task
is to convert it into a JSON object following the exact schema provided
below.

STRICT RULES:

1. Output valid JSON only. Do not include markdown, explanations, or
extra text.

2. Use only the predefined field names in the schema. Do not add new
keys.

3. summary:

   - Copy the source-provided core summary verbatim.
   
   - Do not paraphrase, expand, or rewrite it.
   
   - If the input contains separate summary and detailed-description fields, use only the designated summary field.
     
   - If no explicit core summary is provided, copy the original natural-language description verbatim; do not generate a new summary.

4. speech:

   - If spoken words are explicitly quoted, copy them exactly.
   
   - Otherwise leave it empty.
   
   - Do not create speech content for clips without human speech.
   
5. lyrics: fill only when song lyrics are explicitly provided. Otherwise leave it empty.
   
6. Select exactly one type:

   speech: dominant human speech;
   
   singing: dominant singing or vocal performance;
   
   music: non-vocal instrumental music;
   
   sfx: discrete sound events such as impacts, machines, animals, or weather;
   
   mixed: two or more foreground acoustic categories among speech, singing,
music, and sound effects coexist and are jointly relevant;
   
   unknown: insufficient information.

7. language:

   - Use an ISO language code only when human speech or singing is
   present.
   
   - Otherwise leave it empty.
   
8. sfx should describe specific discrete events rather than generic
labels.

9. ambience describes persistent environmental background.

10. texture describes recording or acoustic characteristics only.

11. vocal describe voice characteristics only.

12. speaker\_count must be one of: "1", "2", "unknown", or "".

13. music should describe only explicitly supported musical information,
including instruments, rhythm, style, or mood when available.

14. Do not infer missing information from world knowledge or context.
All unsupported fields must remain empty.
All field values should be concise noun phrases rather than complete sentences.

JSON Schema (output exactly this structure):

\{
  "summary": "",
  "type": "",
  "lang": "",
  "speech": "",
  "lyrics": "",
  "speaker\_count": "",
  "relation": "",
  "vocal": "",
  "music": "",
  "sfx": "",
  "ambience": "",
  "texture": ""
\}
\end{promptbox}
The extracted JSON representation is deterministically serialized into
the XML-like tagged text format consumed by the text encoder; this
serialization changes only the representation format and does not add
semantic information.
\subsection{Public-Benchmark Prompt Rendering}
\label{app:benchmark_conversion}

\begin{table*}[t]
\centering
\caption{Representative prompts.  The examples illustrate the field semantics; they do not define a required set of fields for every sample.}
\label{tab:prompt-examples}
\begin{tabular}{p{0.15\linewidth}p{0.75\linewidth}}
\toprule
Scenario & Structured prompt \\
\midrule
TTS & \texttt{plain English speech <type>speech</type> <lang>en</lang> <speech>Keep walking along this road, then turn right at the second intersection.</speech> <speaker\_count>1</speaker\_count> <vocal>adult English voice, neutral tone</vocal>} \\
\addlinespace
TTA & \texttt{A train arrives while people talk on a platform <type>sfx</type> <sfx>train arriving, people talking</sfx> <ambience>train station platform with crowd murmur</ambience>} \\
\addlinespace
TTM & \texttt{Warm instrumental piano with slow chords <type>music</type> <music>instrumental piano, slow warm chords</music> <texture>close studio recording</texture>} \\
\addlinespace
Mixed & \texttt{English race-engineer and driver radio dialogue at high speed <type>mixed</type> <lang>en</lang> <speech>Box this lap; the rear temperatures are climbing. / Copy, but the steering is vibrating under braking. / Understood, stay off the inside curb.</speech> <speaker\_count>2</speaker\_count> <relation>pit engineer advising a race driver</relation> <vocal>controlled male engineer and strained male driver through clipped team radio</vocal> <sfx>Engine roar, wind, tires, and radio pops collide</sfx> <ambience>race car at speed with strong wind trackside</ambience> <texture>driver transmission is noisier and more distorted than the engineer response</texture>} \\
% \addlinespace
% Mandarin riverside singing & \texttt{In a riverside park, an adult female singer softly sings in Mandarin <type>mixed</type> <lang>zh</lang> <lyrics>\begin{CJK*}{UTF8}{gbsn}风从桥下慢慢来，带走水面一片白。\end{CJK*} </lyrics> <speaker\_count>1</speaker\_count> <vocal>soft adult female Mandarin singer</vocal> <music>gently strummed acoustic guitar</music> <sfx>wind blows, birdsong, bicycle bells, distant sound of the park</sfx> <ambience>riverside park</ambience> <texture>voice remains foregrounded within a natural outdoor mix</texture>} \\
\bottomrule
\end{tabular}
\end{table*}

For public benchmarks, the structured representation is used as an interface normalization step rather than an additional source of information. The original benchmark input remains the only semantic source, and the conversion procedure does not access reference audio or introduce additional acoustic descriptions.

For TTS benchmarks, transcripts are preserved exactly in the speech field. The converter only adds schema-level information such as the acoustic type,
language identifier, and available speaker metadata. No words, pronunciation details, or speaker characteristics are generated beyond the original input.

For caption-based benchmarks, including AudioCaps, MusicCaps, and Song Describer, the original caption is retained as the summary field. The structured converter maps the caption into the predefined fields and performs schema validation. In particular, speech and lyrics are preserved verbatim when explicitly present, while events, instruments, attributes, or emotions not supported by the source caption are not introduced.

The same structured interface is used for training and inference, but the benchmark conversion is applied only to the publicly available text descriptions. Since the conversion does not observe the target audio, the resulting structured caption contains no information unavailable from the original benchmark input.

% For controlled comparisons on the Complex-Scene benchmark, SonicWeave,
% Dense, and Base-MoE receive the same structured representation derived
% from the natural-language specification. This isolates the effect of
% the conditioning interface and routing mechanism without providing
% additional target information.

\subsection{Prompt Templates for Unified Audio Scenes}

The following templates illustrate the interface used during training and inference.  Bracketed fields are optional and omitted when unsupported.  A mixed scene explicitly contains at least two of speech, music, singing, and sound effects.

\begin{sloppypar}
\small
\noindent\textbf{Speech:} \texttt{summary <type>speech</type> <lang>LANG</lang> <speech>TEXT</speech> <speaker\_count>N</speaker\_count> <vocal>VOICE</vocal> [<ambience>ROOM</ambience>] [<texture>MIX</texture>]}

\noindent\textbf{Music:} \texttt{summary <type>music</type> <music>INSTRUMENTS, STYLE, TEMPO</music> [<texture>PRODUCTION</texture>]}

\noindent\textbf{Singing:} \texttt{summary <type>singing</type> <lang>LANG</lang> <lyrics>LYRICS</lyrics> <vocal>VOCAL STYLE</vocal> <music>ACCOMPANIMENT</music> [<texture>MIX</texture>]}

\noindent\textbf{SFX:} \texttt{summary <type>sfx</type> <sfx>EVENTS</sfx> \\<ambience>ENVIRONMENT</ambience> [<texture>RECORDING CHARACTER</texture>]}

\noindent\textbf{Mixed scene:} \texttt{summary <type>mixed</type> <lang>LANG</lang> <speech>TEXT</speech> <speaker\_count>N</speaker\_count> <vocal>VOICE</vocal> [<music>BED</music>] [<sfx>EVENTS</sfx>] [<ambience>ENVIRONMENT</ambience>] \\ <texture>FOREGROUND--BACKGROUND RELATION</texture>}
\end{sloppypar}
\normalsize

\subsection{Representative Structured Prompts}

Table~\ref{tab:prompt-examples} illustrates how the renderer preserves source information.  For TTS, the original speech content is placed directly in \texttt{speech} without adding scene content.  For TTA and TTM, the original caption is retained verbatim as the first summary sentence, with the remaining fields only reorganizing information already present in that caption.  The two complex-scene rows show how speech or lyrics, music, sound effects, ambience, and their foreground--background relation are represented together.

\section{Training and Inference Details}
\label{app:implementation}

\subsection{Training Setup}

The reported model is trained on approximately five million clips of 10--15 seconds from an internal corpus exceeding 20,000 hours.  The corpus combines speech, singing, music, sound effects, and mixed recordings under the unified prompt interface.  The approximate content
distribution is 20\% speech, 5\% singing, 10\% music, 10\% sound
effects, and 55\% mixed recordings. We use the same data distribution, DiT backbone, text encoder, and training budget for Dense, Base-MoE, and SonicWeave; only the FFN/routing mechanism differs in the controlled comparison.

The final training run used AdamW with a peak learning rate of $7.5\times10^{-5}$.  The learning-rate schedule consists of the configured linear warm-up followed by linear decay.  Training used fp16 mixed precision, with a global batch size of 512 and one gradient-accumulation step, for 800,000 optimizer updates.  The trainer maintained an exponential moving average with decay 0.9999 after a 1,000-step warm-up.  We selected the final checkpoint for evaluation and used its EMA parameters when available; no intermediate checkpoint was selected by a validation-score sweep.

\subsection{Conditioning Dropout and Reference Masking}

For each training sample, a contiguous span whose fractional length is sampled uniformly from $[0.7,1.0]$ is selected from valid latent frames.  This region is zeroed in the reference condition but remains visible in the clean target and flow objective.  The mechanism trains continuation and inpainting behavior without changing the flow target.

Classifier-free guidance conditions are sampled as follows.  The audio reference is independently dropped with probability 0.3 and the caption with probability 0.1.  Caption drop zeros both text tokens and the pooled global text condition.  When audio reference is retained, $\mathcal{L}_{\mathrm{FM}}$ is restricted to the selected masked span; when it is dropped, the loss is evaluated over all valid frames.  This makes the null branch a genuine text-to-audio training condition rather than a masked-reference reconstruction condition.

\subsection{Adaptive Projected Guidance}

At ODE state $\mathbf{x}_t$, let $\mathbf{v}_c$ and
$\mathbf{v}_u$ denote the conditional and unconditional
velocity predictions evaluated at the same state. Under the
linear flow path, their clean-sample estimates are
\begin{equation}
\boldsymbol{\mu}_c
=
\mathbf{x}_t+(1-t)\mathbf{v}_c,
\qquad
\boldsymbol{\mu}_u
=
\mathbf{x}_t+(1-t)\mathbf{v}_u.
\end{equation}
We define
$\Delta\boldsymbol{\mu}
=
\boldsymbol{\mu}_c-\boldsymbol{\mu}_u$
and apply reverse momentum,
\begin{equation}
\widetilde{\Delta\boldsymbol{\mu}}_k
=
\Delta\boldsymbol{\mu}_k
+
\beta\widetilde{\Delta\boldsymbol{\mu}}_{k-1},
\qquad
\beta=-0.1,
\end{equation}
with
$\widetilde{\Delta\boldsymbol{\mu}}_0
=
\Delta\boldsymbol{\mu}_0$.
We decompose the smoothed guidance direction with respect to
the conditional clean estimate:
\begin{equation}
\Delta\boldsymbol{\mu}_{\parallel}
=
\frac{
\langle
\widetilde{\Delta\boldsymbol{\mu}},
\boldsymbol{\mu}_c
\rangle
}{
\|\boldsymbol{\mu}_c\|_2^2+\epsilon
}
\boldsymbol{\mu}_c,
\qquad
\Delta\boldsymbol{\mu}_{\perp}
=
\widetilde{\Delta\boldsymbol{\mu}}
-
\Delta\boldsymbol{\mu}_{\parallel}.
\end{equation}
The guided clean estimate is
\begin{equation}
\boldsymbol{\mu}_{\mathrm{APG}}
=
\boldsymbol{\mu}_u
+
w\left(
\Delta\boldsymbol{\mu}_{\perp}
+
\eta\Delta\boldsymbol{\mu}_{\parallel}
\right),
\end{equation}
and is converted back to velocity space as
\begin{equation}
\mathbf{v}_{\mathrm{APG}}
=
\frac{
\boldsymbol{\mu}_{\mathrm{APG}}-\mathbf{x}_t
}{
\max(1-t,10^{-5})
}.
\end{equation}
% \subsection{Adaptive Projected Guidance}

% At ODE state $\mathbf{x}_t$, let $\mathbf{v}_c$ and $\mathbf{v}_{\varnothing}$ be the conditional and unconditional velocity predictions.  Under the linear flow path, they imply clean-sample estimates
% \begin{equation}
% \boldsymbol{\mu}=\mathbf{x}_t+(1-t)\mathbf{v}_c,\qquad
% \boldsymbol{\mu}_{\varnothing}=\mathbf{x}_{\varnothing}+(1-t)\mathbf{v}_{\varnothing},
% \end{equation}
% where $\mathbf{x}_{\varnothing}$ equals $\mathbf{x}_t$ for pure generation.  Define $\Delta\boldsymbol{\mu}=\boldsymbol{\mu}-\boldsymbol{\mu}_{\varnothing}$.  We use reverse momentum
% \begin{equation}
% \widetilde{\Delta\boldsymbol{\mu}}_k=
% \Delta\boldsymbol{\mu}_k+\beta\widetilde{\Delta\boldsymbol{\mu}}_{k-1},
% \qquad \beta=-0.1,
% \end{equation}
% with the first step initialized by $\Delta\boldsymbol{\mu}_0$.  The direction is decomposed per latent frame into
% \begin{equation}
% \Delta\boldsymbol{\mu}_{\parallel}=
% \frac{\langle\widetilde{\Delta\boldsymbol{\mu}},\boldsymbol{\mu}\rangle}
% {\|\boldsymbol{\mu}\|_2^2+\epsilon}\boldsymbol{\mu},
% \qquad
% \Delta\boldsymbol{\mu}_{\perp}=
% \widetilde{\Delta\boldsymbol{\mu}}-\Delta\boldsymbol{\mu}_{\parallel}.
% \end{equation}
% The APG estimate and velocity are
% \begin{equation}
% \boldsymbol{\mu}_{\mathrm{APG}}=
% \boldsymbol{\mu}+w\left(\Delta\boldsymbol{\mu}_{\perp}+
% \eta\Delta\boldsymbol{\mu}_{\parallel}\right),
% \end{equation}
% \begin{equation}
% \mathbf{v}_{\mathrm{APG}}=
% \frac{\boldsymbol{\mu}_{\mathrm{APG}}-\mathbf{x}_t}{\max(1-t,10^{-5})},
% \end{equation}
where $w$ is the guidance scale. We use $\eta=0.85$ for speech-only generation and $\eta=0.5$ when the requested scene contains sound effects or music; the latter setting gave the best results for these mixed conditions.  APG is disabled when $w<10^{-5}$.  The unconditional pass uses dropped caption and dropped audio reference; for masked-reference editing, the noisy-state masking policy is applied consistently to that pass~\cite{apg}.

\subsection{Sampling Protocol}

Unless a task-specific protocol states otherwise, SonicWeave uses Euler integration over 100 uniform time points from 0 to 1, fixed random seed 42, no reference audio, and full-span latent generation.  The parallel inference entry point uses guidance scale 4.5 for the reported generation path.  The model supports a cosine sway schedule at lower sampling budgets, but the public-benchmark and Complex-Scene results should be generated with the fixed protocol above rather than selecting a schedule per example.

The audio waveform sample rate is 44.1~kHz, and the stereo VAE has a temporal downsampling factor of 2048 samples per latent frame.  We use a pretrained stereo VAE based on the Stable Audio 2.0 VAE architecture~\cite{stableaudio}.  Public-benchmark decoding uses the duration requested by the corresponding benchmark protocol: generated latent length is set from the benchmark item or its task-specific maximum rather than silently truncating all tasks to one common duration.  The model defaults to generate 256 latent frames, corresponding to approximately 11.89s at 44.1~kHz.  Because the flow-matching DiT operates on a variable-length latent sequence, it is not architecturally restricted to this default duration.

\section{Evaluation Protocols}
\label{app:evaluation}

\subsection{Public-Benchmark Protocols}

We evaluate text-to-speech (TTS) on SeedTTS-eval and LibriSpeech-PC test-clean, text-to-audio (TTA) on AudioCaps, and text-to-music (TTM) on MusicCaps and Song Describer.  For TTS, we report word error rate (WER) on English benchmarks and character error rate (CER) on Mandarin benchmarks.  TTA is evaluated using VGGish-based Fr'echet distance ($\mathrm{FD}_{\mathrm{VGG}}$, also referred to as FAD), PaSST-based KL divergence, and CLAP text--audio similarity.  TTM is evaluated using PaSST-based KL divergence and CLAP similarity.  Lower WER/CER, $\mathrm{FD}_{\mathrm{VGG}}$, and KL indicate better performance, whereas higher CLAP is better.

Each system generates one audio sample per evaluation item under a common fixed-seed policy.  WER/CER is computed between the benchmark transcript and an ASR transcription of the generated audio after applying the official evaluation normalization, including punctuation removal and English lowercasing.  CLAP is computed as the mean cosine similarity between the text-prompt and generated-audio embeddings.  $\mathrm{FD}_{\mathrm{VGG}}$ and KL are aggregate metrics and are reported once for each benchmark set.  The number of retained items is obtained from the final evaluation manifest for each benchmark.

For TTS transcription, we follow the public SeedTTS/F5-TTS evaluation pipeline, using faster-Whisper large-v3~\cite{whisper} for English and FunASR Paraformer-zh~\cite{paraformer} for Mandarin.  For AudioCaps, $\mathrm{FD}_{\mathrm{VGG}}$ is computed using the default VGGish-based implementation in AVBench.  PaSST KL follows the Stable Audio Metrics implementation.  CLAP is evaluated with LAION-CLAP, using the general-audio checkpoint \small\texttt{630k-audioset-fusion-best.pt} \normalsize for TTA and \small\texttt{music\_speech\_audioset\_epoch\_15\_esc\_89.98.pt} \normalsize for TTM. 

\subsection{Complex-Scene Suite}

\begin{table*}[t]
\centering
\caption{
Targeted routing ablations on representative benchmarks.
All variants share the same backbone, training data, expert
configuration, and training budget. Complex-scene results use
the same reference-free automatic judging protocol as the main
evaluation.
}
\label{tab:routing_ablation}
\setlength{\tabcolsep}{4pt}
\begin{tabular}{lccccccccc}
\toprule
Variant
& \multicolumn{2}{c}{SeedTTS}
& \multicolumn{3}{c}{AudioCaps}
& \multicolumn{2}{c}{Song Describer}
& \multicolumn{2}{c}{Complex Scene} \\
\cmidrule(lr){2-3}
\cmidrule(lr){4-6}
\cmidrule(lr){7-8}
\cmidrule(lr){9-10}
& En WER$\downarrow$
& Zh CER$\downarrow$
& FAD$\downarrow$
& KL$\downarrow$
& CLAP$\uparrow$
& KL$\downarrow$
& CLAP$\uparrow$
& AI-Tech$\uparrow$
& AI-Sem$\uparrow$ \\
\midrule
Fixed gate ($g_j=0.5$, $C=4$)
& 2.1\% & 1.8\% & 3.69 & 1.54 & 0.448 & 0.53 & 0.31 & 4.60 & 4.35\\
CPE-MoE ($C=8$)
& 1.4\% & 0.9\% & 2.91 & 1.29 & 0.469 & 0.44 & 0.33 & 4.69 & 4.64 \\
CPE-MoE ($C=4$)
& \textbf{1.0\%} & \textbf{0.8\%} & \textbf{2.75} & \textbf{1.26} & \textbf{0.475} & \textbf{0.36} & \textbf{0.44} & \textbf{4.79} & \textbf{4.72}\\
\bottomrule
\end{tabular}
\end{table*}

The Complex-Scene suite contains 100 prompts across four categories: 45 speech-with-background cases, 10 singing-with-background cases, 15 clean dialogue cases, and 30 dialogue-with-background cases.  It tests not only whether individual components occur, but whether linguistic content, speaker structure, music, sound effects, ambience, and foreground--background relationships are jointly realized.  The suite includes Mandarin and English prompts, with clean dialogue serving as a control condition and the remaining categories emphasizing coexistence, masking, and composition.

Each case has a natural-language prompt and a paired structured prompt.  The natural-language form specifies the complete target scene in ordinary prose.  The structured form reorganizes the same specification into the fields of Table~\ref{tab:prompt-schema}; it does not add new target events.  Public systems receive the natural-language form in their native interface.  SonicWeave, Dense, and Base-MoE receive the paired structured form.  This setup evaluates structured conditioning and CPE-MoE under the same requested semantics while avoiding unsupported claims of identical tokenizer-level inputs across unrelated public systems.

Representative cases are listed in Table~\ref{tab:prompt-examples}.  They include speech--background, singing--background, clean dialogue, and dialogue--background conditions, with both continuous background layers and short high-salience events.  The suite therefore tests whether a system preserves foreground speech or singing while realizing local acoustic departures and their intended foreground--background relation.

\subsection{Reference-Free Judge Rubric}

We use \texttt{gemini-3.1-pro-preview} as a blind audio judge when a reference recording is unavailable.  The judge receives one generated audio clip, the natural-language prompt, and the paired structured specification.  Its instruction is: ``Listen to the audio first; score only audible evidence against the requested scene; do not credit an event solely because it appears in the prompt; return the required JSON fields.'' The system/model name is not supplied.

Each applicable dimension is scored on a five-point ordinal scale represented internally as $1$--$5$: 1 denotes absent, contradictory, or completely wrong; 2 mostly wrong with limited relevant evidence; 3 partially correct with substantial omissions or errors; 4 mostly correct with minor omissions; and 5 fully and accurately realized.  The seven dimensions are:
\begin{itemize}
\item \textbf{Speech content}: required words and language;
\item \textbf{Vocal attributes}: speaker type, emotion, energy, delivery, and style;
\item \textbf{Speaker structure}: count, identity separation, turns, and assignment;
\item \textbf{Background adherence}: requested music, events, and ambience;
\item \textbf{Acoustic relation}: prominence, masking, timing, overlap, and interaction;
\item \textbf{Scene coherence}: whether components form a plausible integrated scene; and
\item \textbf{Technical quality}: clipping, distortion, discontinuity, abrupt cuts, and synthesis artifacts.
\end{itemize}
For clean dialogue, background-adherence and acoustic-relation are marked not applicable rather than treated as zeros.  The semantic score is the mean of all applicable first six dimensions; technical quality is retained separately. 

The judge must return a short audible-evidence string for every score, detected languages, detected speaker count, an ordered transcript where applicable, detected background elements, and a confidence score.  Outputs failing the required JSON schema or score ranges are rejected and retried.  Temperature is zero to reduce sampling variation.

\subsection{Human Listening Study}

The listening study presents every system output anonymously.  For each case, available clips are assigned anonymous identifiers and their order is randomized with a fixed study seed.  The private identifier-to-system mapping is not shared with listeners.  Listeners score each clip independently on two 1--5 scales:
\begin{itemize}
\item \textbf{MOS-Q} (quality): 1 is nearly unusable due to severe noise, clipping, distortion, discontinuity, or unintelligibility; 2 has major persistent artifacts that substantially impair listening; 3 is basically listenable with noticeable artifacts; 4 is clear and stable with only minor artifacts; 5 is clear, natural, stable, and nearly artifact-free.
\item \textbf{MOS-R} (request realization): 1 means that the requested scene is essentially absent or contradictory; 2 means that only isolated requested elements are recognizable while key content or relations are missing; 3 means major requested content is present but with clear omissions or incorrect relationships; 4 means nearly all requested content and relations are realized with minor omissions; 5 means foreground speech, music, events, ambience, timing, and relative prominence are jointly and naturally realized.
\end{itemize}

The listening study used 25 listener responses.  Every listener received the same questionnaire containing 25 cases and six anonymously shuffled system outputs per case, for 150 clips and 300 ordinal ratings (MOS-Q and MOS-R) per completed response.  The study script randomizes clip order within each case with seed 42, hides the system mapping, and provides written 1--5 scoring instructions.  The aggregation excludes ratings with missing mappings or values outside $[1,5]$. MOS results are reported as mean ± standard deviation.

\section{Targeted Routing Ablations}
\label{app:ablations}

\subsection{Scope and Evaluation Protocol}

The main paper compares SonicWeave with matched Dense and Base-MoE controls.  Targeted ablations isolate two specific CPE-MoE design choices rather than exhaustively decomposing all architectural factors.  Every ablation retains the same data, backbone width/depth, number of routed layers, expert count, top-$K$, optimizer schedule, training budget, prompt interface, and sampling protocol unless the row explicitly changes one variable.

All ablation variants are evaluated with objective metrics across TTS, TTA, TTM, and Complex-Scene reference-free judging.  We report English WER and Mandarin CER, AudioCaps FAD/KL/CLAP, Song Describer KL/CLAP, and Complex-Scene AI-Tech/AI-Sem.  Human MOS is reserved for the main model comparison and is not used to rank targeted ablations.

\subsection{Fixed Prior--Evidence Fusion}
\label{app:fixed_gate}

The fixed-fusion control replaces the learned chunk gate by $g_j=0.5$ for every valid chunk, layer, prompt, and diffusion phase:
\begin{equation}
\boldsymbol{\ell}_j=0.5\boldsymbol{\ell}^{\mathrm{prior}}+0.5\boldsymbol{\ell}^{\mathrm{evid}}_j.
\end{equation}
It retains $C=4$ and all remaining CPE-MoE components.  This control tests whether adaptive, content- and phase-dependent fusion improves over a uniform average.  It does \emph{not} separately identify the causal contribution of the prior branch or evidence branch, because both are still present.  The reported table shows that this control is consistently worse than the learned gate: English WER and Chinese CER increase, AudioCaps FAD/KL/CLAP degrade, Song Describer KL/CLAP degrade, and Complex-Scene AI-Tech/AI-Sem decrease.  The degradation is especially clear on TTS, which is consistent with a uniform fusion rule limiting the adaptive balance between the text-derived prior and evolving acoustic evidence.

\subsection{Chunk-Size Sensitivity}
\label{app:chunk-sensitivity}

The $C=8$ variant is broadly comparable to $C=4$ on public benchmarks, but is consistently worse on every displayed TTS, AudioCaps, and Song Describer metric.  The difference is modest for globally coherent content, such as sustained music, because a longer chunk can still share a suitable expert path.  However, each $C=8$ assignment constrains twice as many neighboring latent frames as $C=4$.  This reduces the routing frequency available to react to short events, speaker turns, or local changes in foreground--background balance.  
The effect is clearest on Complex-Scene, where linguistic content and transient acoustic events must be jointly realized.  
Conversely, smaller chunks would increase routing overhead and can fragment a locally coherent acoustic event.  We therefore use $C=4$ as the trade-off between local adaptivity, short-range routing consistency, and sparse-dispatch efficiency.

\section{Additional Routing Analyses}
\label{app:analysis}

\begin{figure}[t]
    \centering
    \includegraphics[width=\linewidth]{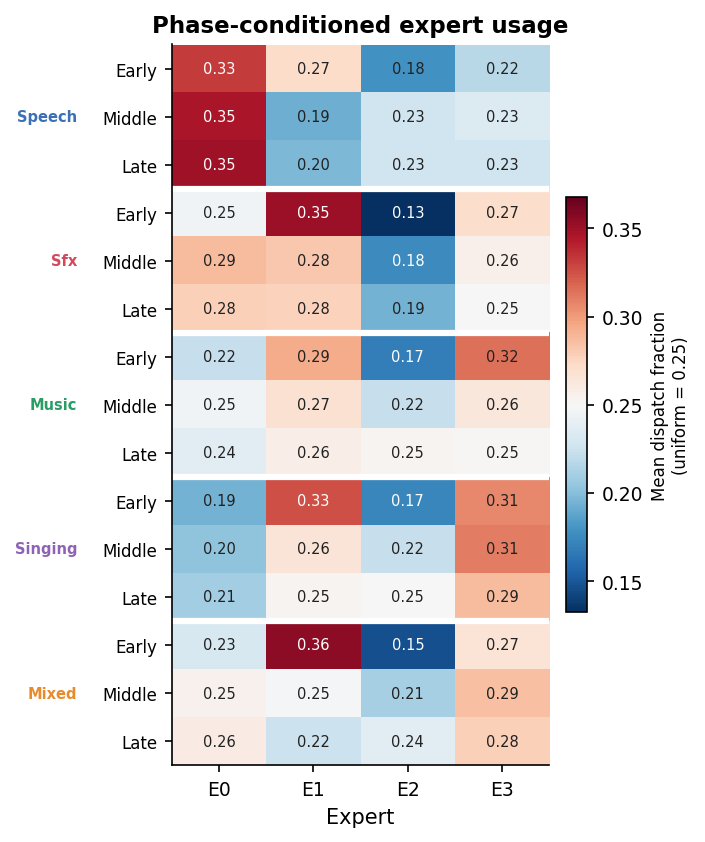}
    \caption{\textbf{Phase- and content-conditioned expert usage.} Mean dispatch fractions are shown for each content type over the early, middle, and late thirds of the diffusion trajectory.  Content-specific preferences persist across phases, while the degree of concentration changes as generation proceeds.}
    \label{fig:phase-expert-usage}
\end{figure}

\subsection{Routing-Analysis Protocol}

We construct a balanced 250-prompt routing benchmark with 50 prompts each for speech, sound effects, music, singing, and mixed scenes.  Speech is balanced across SeedTTS English and Chinese examples; sound effects use AudioCaps; music and singing use MusicCaps; mixed examples use prompts from complex-scene test sets.  Each item is assigned a fixed prompt identifier and a maximum frame length appropriate to its source, preventing a content type from being overrepresented solely through longer sequences.

For every routed layer, generation step, and valid chunk, we record the conflict gate, posterior routing distribution, and top-$K$ dispatch.  All aggregate statistics are first averaged within each prompt and then across prompts, so that long clips or clips containing more valid chunks do not receive disproportionate weight.
% These analyses characterize the learned routing mechanism; generation quality is assessed separately by the matched objective and listening evaluations reported in the main paper.

\subsection{Phase- and Content-Conditioned Expert Usage}

\begin{figure*}[t]
    \centering
    \includegraphics[width=0.8\textwidth]{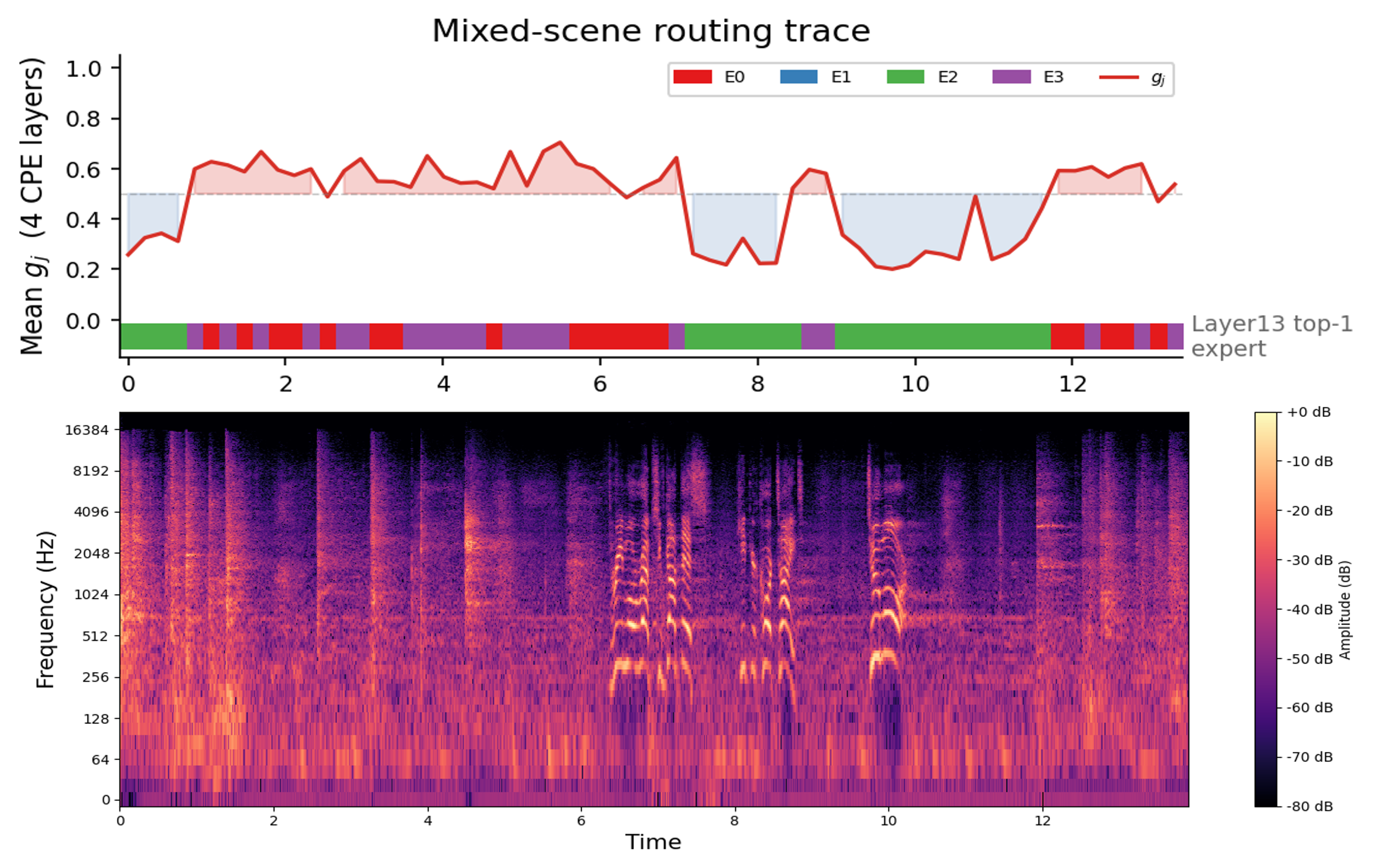}
   \caption{\textbf{Conflict-gate dynamics and mixed-scene routing.}  The gym scene \texttt{cpe\_010}, showing the gate averaged over four CPE layers and Layer~13 top-1 expert assignments. Complex background regions rely more on acoustic evidence, while the main speech segment shows stronger reliance on the text prior.}
    \label{fig:mixed-routing-case}
\end{figure*}

Figure~\ref{fig:phase-expert-usage} reports mean expert dispatch fractions for each content type during the early, middle, and late thirds of the generation trajectory.  The dispatch patterns depend jointly on content and phase.  Speech retains a strong preference for E0, while singing consistently uses E3 more heavily.  Sound effects, music, and mixed scenes exhibit sharper preferences early in generation and become more balanced as the sample is refined.  Thus, the routed experts do not form a single static partition of the training domains: their relative utilization evolves during generation while preserving persistent content-dependent tendencies.

\subsection{Does the Gate Control the Effective Router?}
\begin{figure}[h]
    \centering
    \includegraphics[width=\linewidth]{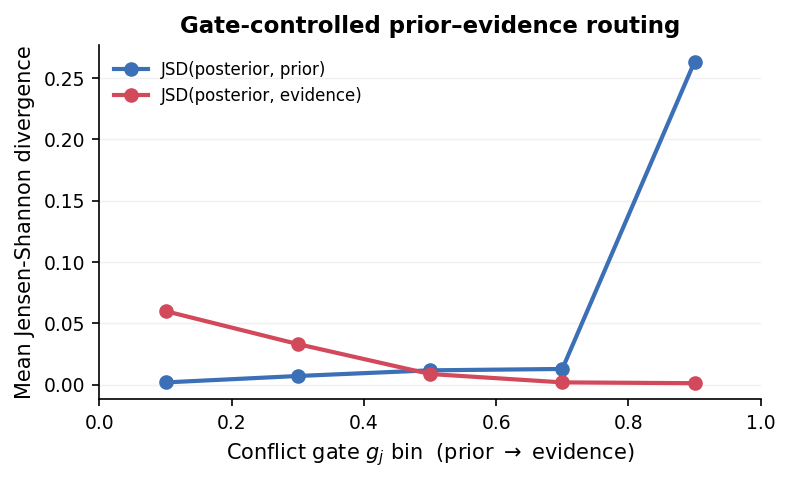}
    \caption{\textbf{Gate-controlled prior--evidence routing.} Valid chunks are grouped by their conflict-gate values.  At low $g_j$, the fused posterior is close to the global prior; at high $g_j$, it is close to local evidence.  JSD denotes Jensen--Shannon divergence.}
    \label{fig:gate-routing-coupling}
\end{figure}
The conflict gate is designed to interpolate the prior and evidence logits, but this definition alone does not show whether trained gates occupy a useful operating range or materially alter the resulting routing distribution.  We therefore bin valid chunks by $g_j$ and measure the Jensen--Shannon divergence (JSD) between the fused posterior distribution and each of its two inputs.  As shown in Figure~\ref{fig:gate-routing-coupling}, low-gate chunks remain close to the global prior and comparatively far from local evidence.  The relation reverses for high-gate chunks: the posterior approaches the evidence distribution and departs sharply from the prior.  Intermediate bins provide a smooth transition between these regimes.  This aggregate behavior confirms that the learned gate controls the effective routing distribution according to its intended prior-to-evidence semantics.

\subsection{Representative Mixed-Scene Routing Case}

Figure~\ref{fig:mixed-routing-case} presents a representative case from the demo set (prompt identifier \texttt{cpe\_010: ``In a busy gym, a trainer shouts, `Three more reps, you've got this! Keep your core tight. Two more!' Weight plates clank, treadmills run, people breathe heavily, and battle ropes strike the floor around him.''})  The scene combines foreground speech with continuous machine and breathing noise, together with transient or quasi-periodic impacts.  It therefore provides a natural test of within-clip routing under overlapping heterogeneous content.

The upper trace shows the conflict gate averaged over the four CPE-MoE layers, whereas the colored strip visualizes top-1 dispatch from a single routed layer (Layer~13). We include one layer to keep the visualization readable; the layer is used only as an illustrative dispatch trace and is not intended to summarize routing behavior across all routed layers.

The trace shows a coarse but interpretable correspondence between the acoustic composition and the routing behavior. During the earlier portion of the clip, the scene is dominated by dense background activity, including machine noise, breathing, and transient impacts. In this region, the mean gate is mostly above 0.5, indicating stronger reliance on the evolving acoustic evidence. The Layer~13 dispatch trace also switches experts more frequently around local acoustic changes, consistent with chunk-level adaptation to heterogeneous and transient content.

Around 7 seconds, the main speech segment becomes prominent in both the spectrogram and the corresponding demo audio. This transition coincides with a sustained decrease of the mean gate below 0.5, shifting the routing toward the structured text prior, which provides a stable description of the foreground speech. The expert assignments simultaneously enter longer, more stable runs. Near the end of the clip, as the background activity becomes more prominent again, the gate rises and the expert pattern changes accordingly.

Although the correspondence is not frame-exact, the case illustrates the intended behavior of CPE-MoE: local acoustic evidence plays a larger role in complex and rapidly changing regions, whereas the global text prior becomes more influential during semantically well-specified speech segments.

% The trace exhibits extended routing regimes rather than frame-level fluctuations.  The earlier portion is predominantly evidence-weighted ($g_j>0.5$), whereas a later speech-dominant portion audible in the corresponding demo coincides with a sustained lower-gate interval and a different top-1 expert pattern.  We use \emph{coincides with} deliberately: the prompt does not provide frame-level event boundaries, so the visualization supports a qualitative alignment between acoustic transitions and routing behavior rather than a causal attribution of individual gate changes to named events.  Taken together with the population-level analyses in Figures~\ref{fig:phase-expert-usage} and~\ref{fig:gate-routing-coupling}, this case illustrates how CPE-MoE can reorganize routing locally within a single mixed scene while maintaining chunk-level consistency.

% \bibliographystyle{plainnat}

\end{document}